\documentclass[twocolumn,floatfix,letterpaper,prc]{revtex4}

\usepackage{citesort,enumerate}
\usepackage{amsmath,amssymb,amsfonts} % AMS math & symbols
\usepackage{bm}              % bold math
\usepackage{amscd}           % for extensible arrows (e.g., limits)
\usepackage{graphicx}        % PostScript figures
\usepackage{hhline,multirow} % for nicer tables
\usepackage{dcolumn}         % Align table columns on decimal point

\newcommand{\beq}{\begin{equation}}
\newcommand{\eeq}{\end{equation}}
\newcommand{\bea}{\begin{eqnarray}}
\newcommand{\beas}{\begin{eqnarray*}}
\newcommand{\beau}[1]{\begin{equation} \label{#1} \begin{array}{rcl}}
\newcommand{\eea}{\end{eqnarray}}
\newcommand{\eeas}{\end{eqnarray*}}
\newcommand{\eeau}{\end{array} \end{equation}}
\newcommand{\bay}{\begin{array}}
\newcommand{\eay}{\end{array}}
\newcommand{\bals}{\begin{align*}}
\newcommand{\eals}{\end{align*}}

\newcommand{\ra}{{\rightarrow}}

\newcommand{\vev}[1]{\langle #1 \rangle}

\newcommand{\OO}{{\mathcal O}}
\newcommand{\PP}{{\mathcal P}}

\newcommand{\qhat}{{\hat q}}

\newcommand{\Etrf}{\!{E_e^\text{\it trf}}}

\begin{document}

%%%%%%%%%%%%%%%%%%%%%%%%%%%%%%%%%%%%%%%%%%%%%%%%%%%%%%%%%%%%%%%%%%%%%%
%%%%%%%%%%%%%%%%%%%%%%%   TITLE    %%%%%%%%%%%%%%%%%%%%%%%%%%%%%%%%%%%
%%%%%%%%%%%%%%%%%%%%%%%%%%%%%%%%%%%%%%%%%%%%%%%%%%%%%%%%%%%%%%%%%%%%%%

%\preprint{preprint/number}

\title{Final state interactions and hadron quenching in cold nuclear matter}

\author{A.~Accardi}
\affiliation{Dept. of Physics and Astronomy, Iowa State U., Ames, IA
  50011, USA}
%\date{May 25th, 2007}

\begin{abstract}
I examine the role of final state interactions in
cold nuclear matter in modifying hadron production on nuclear targets
with leptonic or hadronic beams. I demonstrate the extent to
which available experimental data in electron-nucleus collisions 
can give direct information on final state effects in hadron-nucleus and
nucleus-nucleus collisions. For hadron-nucleus collisions, a theoretical
estimate based on a parton energy loss model tested in lepton-nucleus
collisions shows a large effect on mid-rapidity hadrons at fixed
target experiments. At RHIC energy, the effect is large for negative
rapidity hadrons, but mild at  
midrapidity. This final state cold hadron quenching
needs to be taken into account in jet tomographic analysis of the
medium created in nucleus-nucleus collisions.
\end{abstract}

%\pacs{...........}

%\keywords{........} 

\maketitle

%%%%%%%%%%%%%%%%%%%%%%%%%%%%%%%%%%%%%%%%%%%%%%%%%%%%%%%%%%%%%%%%%%%%%%
%%%%%%%%%%%%%%%%%%%%%%%   PAPER    %%%%%%%%%%%%%%%%%%%%%%%%%%%%%%%%%%%
%%%%%%%%%%%%%%%%%%%%%%%%%%%%%%%%%%%%%%%%%%%%%%%%%%%%%%%%%%%%%%%%%%%%%%

\section{Introduction}
\label{sec:intro}

Hadron production on nuclear targets is strongly influenced by the
presence of cold and hot nuclear matter. The most spectacular effects
is jet quenching in nucleus-nucleus ($A+A$) collisions
\cite{Arsene:2004fa,Back:2004je,Adams:2005dq,Adcox:2004mh}, namely,
the suppression of 
hadron production at large transverse momentum compared to a suitably
scaled cross section in proton-proton collisions. 
This phenomenon is widely used as
a tool to explore the properties of the hot QCD medium produced in
heavy-ion collisions
\cite{Gyulassy:2003mc,Kovner:2003zj,Vitev:2004bh}, 
and as evidence for the creation of a novel state of matter at the Relativistic
Heavy Ion Collider (RHIC)
\cite{Arsene:2004fa,Back:2004je,Adams:2005dq,Adcox:2004mh}, 
possibly the Quark-Gluon Plasma (QGP) \cite{Gyulassy:2004zy}.
Hadron suppression has also been observed in lepton-nucleus ($\ell+A$) 
\cite{Osborne:1978ai,Ashman:1991cx,Airapetian:2000ks,Airapetian:2003mi,
Airapetian:2007vu,Hafidi:2006ig} and hadron-nucleus ($h+A$) collisions 
\cite{Arsene:2004ux,Adams:2006uz,Back:2004bq}, 
where one does not expect the formation of an extended hot medium.
In this case, the target nucleus itself (cold nuclear matter) induces 
the observed suppression of hadron production, which I will refer to as
``cold hadron quenching'', or cold quenching in short.

Nuclear effects in cold nuclear matter can be classified as initial-state or
final-state depending on whether they happen before or after the hard
collision which generates the hard probe. 
Final state (FS) effects can be isolated in semi-inclusive hadron
production in $\ell+A$ collisions. Hadron suppression in these collisions
\cite{Osborne:1978ai,Ashman:1991cx,Airapetian:2000ks,Airapetian:2003mi,
Airapetian:2007vu} is typically attributed to radiative 
energy loss of the struck quark or to nuclear absorption of a colorless
prehadron, see \cite{Accardi:2006ea} and references therein. Initial
state (IS) effects can be experimentally isolated in Drell-Yan processes in
$h+A$ collisions \cite{Badier:1981ci,Alde:1990im,Vasilev:1999fa}, 
and they are attributed to nuclear 
shadowing or radiative energy loss of the incoming parton
\cite{Johnson:2001xf,Arleo:2002ph,Johnson:2006wi}.
In large-$p_T$ hadron production in $h+A$ collisions both IS and FS
effects are present, and they cannot be easily disentangled. They give
rise to a host of interesting effects. They modify the shape of
midrapidity hadron $p_T$ spectra, slightly suppressing it at small
$p_T \lesssim 1-2$ GeV 
and enhancing it at intermediate 2 GeV $\lesssim p_T \lesssim 6$
GeV; this is known as Cronin effect
\cite{Cronin:1974zm,Antreasyan:1978cw,Accardi:2002ik}. 
An extensive study of the rapidity dependence of the Cronin effect
in deuteron-gold ($d+Au$) collisions at RHIC has shown an interesting
evolution of the Cronin effect: the Cronin enhancement grows at
backward rapidity $y-y_{cm} < 0$
\cite{Adams:2004dv,Adler:2004eh,Abelev:2006pp,Abelev:2007nt}; however, at
forward rapidity $y-y_{cm}>0$, the spectrum is suppressed up to large
$p_T$  \cite{Arsene:2004ux,Adams:2006uz,Back:2004bq}, a trend
confirmed also at lower energy  
collisions \cite{Alber:1997sn,Vitev:2006bi}.  
The forward rapidity suppression has been widely interpreted as
evidence for the onset of the Color Glass Condensate, a universal
state of highly saturated quark and gluons in the nuclear wave
function \cite{Kharzeev:2004yx,Jalilian-Marian:2005jf}. However,
explanations in terms of IS energy loss 
and higher-twist shadowing \cite{Vitev:2006bi}, leading-twist shadowing
\cite{Vogt:2004cs}, Sudakov suppression
\cite{Kopeliovich:2005ym} or FS parton recombination \cite{Hwa:2004in}  
have been equally
successful in describing the data. The rise of the Cronin effect at
backward rapidity is more difficult to understand
\cite{Accardi:2004fi}. Explanations in terms of IS anti-shadowing
\cite{Barnafoldi:2005rb} or saturation
\cite{Adams:2004dv} have been proposed.
Finally, recent PHENIX data on neutral pion production in $d+Au$
collisions at midrapidity suggest a small suppression of midrapidty
$\pi^0$ at $p_T\gtrsim 10$ GeV \cite{Adler:2006wg}. It cannot be
explained by the EMC effect, which is effective at $p_T
\gtrsim 15 $ GeV \cite{Eskola:2002kv,Cole:2007ru}, but may accommodate
a small final state energy loss of order 10\%  \cite{Cole:2007ru}.

A consistent interpretation of this wealth of experimental data
requires a deep understanding of IS and FS interactions at the parton
and hadron level, and the development of a unified computational
framework \cite{Vitev:2007ve}. As a contribution to this program, in
this paper I will analyze phenomenologically the contribution of final
state interactions to hadron production in $h+A$ and $A+A$ collisions,
and I will show that it is indeed 
non negligible in the whole backward rapidity hemisphere up to RHIC
energy. At the Large Hadron Collider (LHC) it will be important 
only at very backward rapidity $y-y_{cm}\lesssim 3$. 
In Section~\ref{sec:LO}, I will review the kinematics of 
hadron production at leading order (LO) in perturbative QCD for Deep
Inelastic Scattering (DIS) and nucleon-nucleon collisions (NN), and  
build a dictionary that relates the kinematic variables used in the
2 cases. In Section~\ref{sec:phasespaces}, I will compare the NN and
DIS phase spaces at present and future experiments in terms of either
set of variables, to understand in detail the relevance of NN to DIS
and viceversa. 
In Section~\ref{sec:coldjetquenching}, I will use the developed 
kinematic dictionary to show the extent to which 
present $\ell+A$ experimental data can give information
on final state cold nuclear matter effects in $h+A$ and $A+A$
collisions (collectively, $A+B$ collisions). Then, I will use an energy
loss model tuned to $\ell+A$ data to estimate cold quenching in $h+A$
collisions for midrapidity hadrons at the Super Proton 
Synchrotron (SPS) and at Fermilab (FNAL), and at various rapidities at
RHIC. A comparison of FS and IS effects will be discussed in
Section~\ref{sec:ISvsFS}, and my conclusions reported in
Section~\ref{sec:conclusions}.

\section{Parton production in DIS and NN collisions}
\label{sec:LO}

Considering parton and hadron production at LO in
NN collisions and DIS collisions, it is easy to provide an explicit  
dictionary translating between the variables traditionally used in the
analysis of the 2 processes. 
I will start by discussing the kinematics of parton and hadron production 
in NN collisions in the center-of-mass frame (c.m.f.). I will then
suitably identify the DIS kinematics in terms of NN variables, and 
derive the dictionary. I will work in the framework of
collinear factorization in pQCD, and 
use light-cone coordinates throughout this
discussion: for any 4-vector $a^\mu$ I write $a^\mu=(a^+,a^-,\vec
a_T)$,  where $a^\pm = (a^0\pm a^3)/\sqrt 2$ are the plus- and
minus-momenta and $\vec a_T = (a^1,a^2)$ the transverse momentum. 

\begin{figure}[t]
  \centering
  \includegraphics[width=6cm]{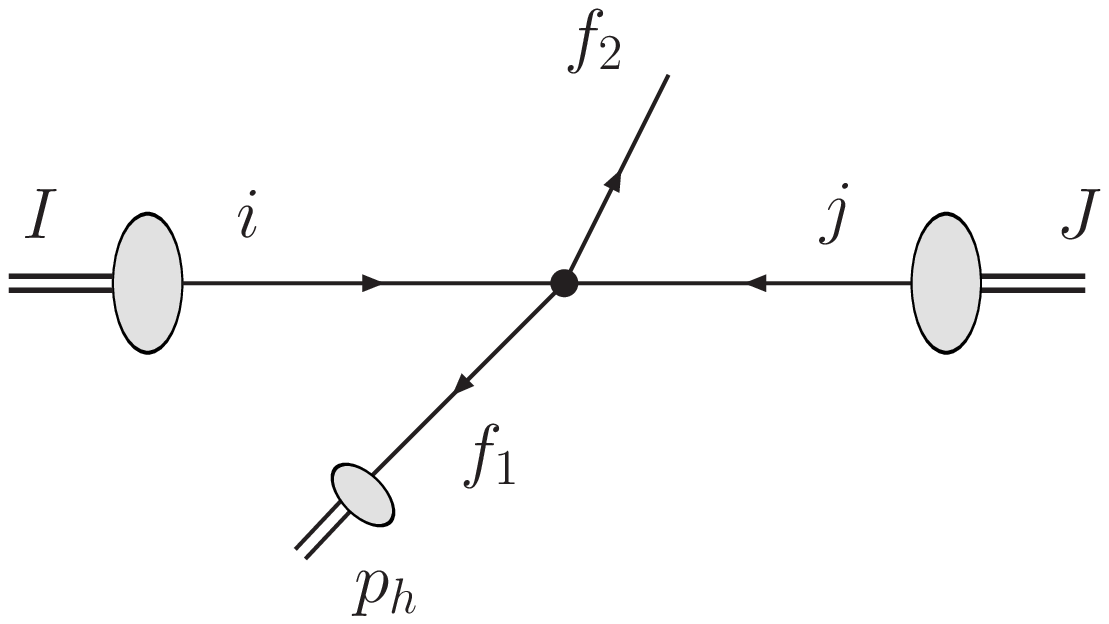}
  \hspace*{0.5cm}
  \includegraphics[width=5.8cm]{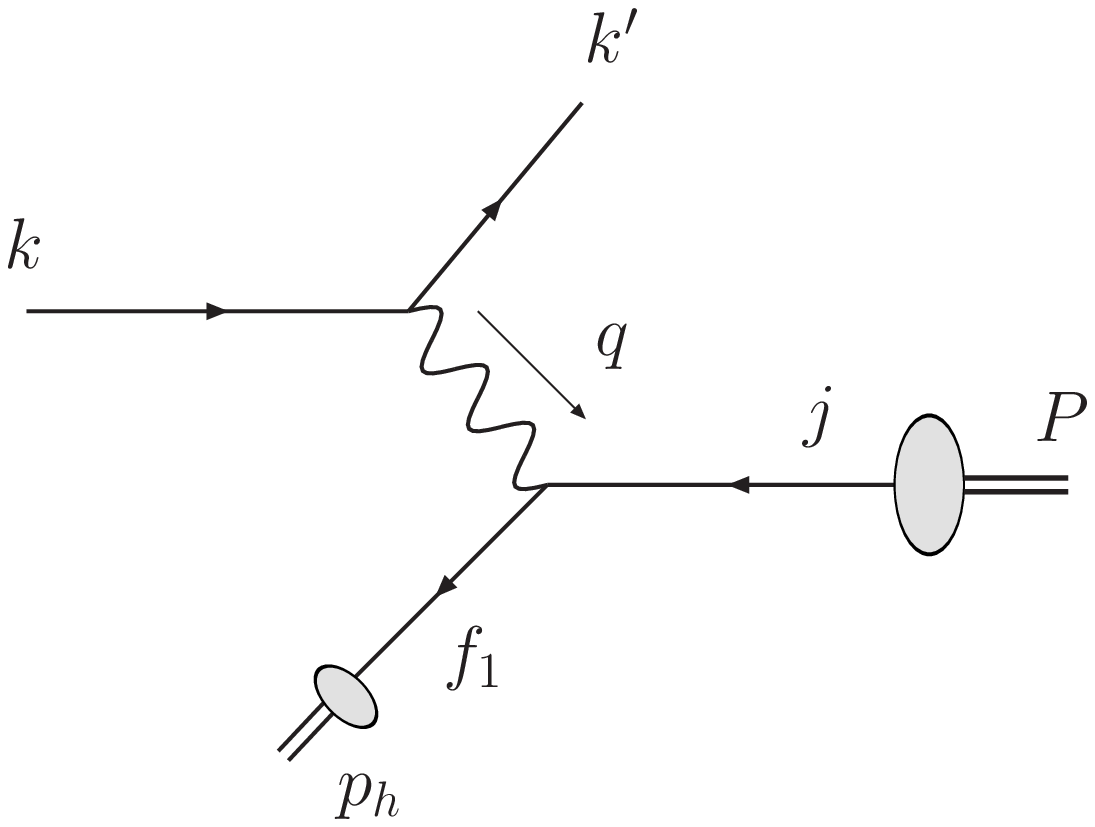}
  \caption{
    {\it Left:} LO kinematics for parton production in NN collisions -- 
    double lines indicate hadrons or nuclei, single lines are partons. 
    {\it Right:} LO kinematics for parton production in DIS collisions -- 
    double lines indicate hadrons or nuclei,
    thin single lines are partons (bottom) or leptons (top). 
    The labels define the particles 4-momenta. }
  \label{fig:NNDISkinematics}
\end{figure}

\begin{table*}[t]
  \begin{tabular}{rcll} 
    \hline 
    {\it Variable} & & {\it Definition} \\\hline
    $\boldsymbol{s}$ & & &  
      \parbox[t]{8.9cm}{\raggedright Nucleon-nucleon center of mass
        energy squared.}\\
    $x_1$ & = & $i^+/I^+$ &  
      \parbox[t]{8.9cm}{\raggedright Initial state projectile parton 
        fractional momentum.}\\
    $x_2$ & = & $j^-/J^-$ &  
      \parbox[t]{8.9cm}{\raggedright Initial state target parton 
        fractional momentum.}\\
    $\vec p_{iT}$ & = & $|\vec{f}_{iT}|$ &
      \parbox[t]{8.9cm}{\raggedright Final state partons transverse 
        momentum (relative to beam).}\\
    $y_i$ & = & $ 0.5 \log(f_i^+/f_i^-)$ &  
      \parbox[t]{8.9cm}{\raggedright Final state partons rapidity.}
      \\ 
    $y_{cm}$ & = & $ 0.5 \log\big(\frac{I^++J^+}{I^-+J^-}\big)$ &  
      \parbox[t]{8.9cm}{\raggedright Rapidity of the center of mass.}
      \\ \hline
    $z$ & = & $p_h^+/f_1^+$ &  
      \parbox[t]{8.9cm}{\raggedright Hadron fractional momentum
        relative to parent parton $f_1$.}\\
    $\boldsymbol{p_{hT}}$ & = & $|\vec{p}_{hT}|$ &  
      \parbox[t]{8.9cm}{\raggedright Hadron transverse momentum
        (relative to beam).}\\ 
    $\boldsymbol{y_h}$ & = & $0.5 \log(p_h^+/p_h^-)$ &  
      \parbox[t]{8.9cm}{\raggedright Hadron rapidity.}\\
    $\boldsymbol{\eta}$ & = & $ - \log\tan(\theta^*/2)$ &  
      \parbox[t]{8.9cm}{\raggedright Hadron pseudorapidity 
        ($\theta^*$ is the angle between the parton and the beam 
        in the center of mass reference frame).}
    \\\hline
  \end{tabular}\\
  \caption{Definitions of the kinematic variables for semi-inclusive
    parton and hadron production in pQCD (top and bottom part of the
    table, respectively). Particle 4-momenta are defined in
    Fig.~\ref{fig:NNDISkinematics}. 
    Boldface variables are experimentally
    measurable. The remaining variables are theoretically defined in
    the QCD parton model in collinear factorization. Note that at LO,
    with 2 final state partons, $\vec p_{1T} = -\vec p_{2T} = \vec p_T$.} 
  \label{tab:NNkvar}
\end{table*}

\begin{table*}[t]
  \begin{tabular}{rclcll} 
    \hline 
    \it Variable & \multicolumn{2}{l}{\it Definition} & 
      \multicolumn{2}{l}{\it Target rest frame} & \\\hline
    \\[-.3cm]
    $\boldsymbol{M^2}$ & = & $P^2$ & & 
      & \parbox[t]{10.7cm}{\raggedright Target mass.}\\
    $\boldsymbol{x_B}$ & = & $\frac{-q^2}{2P\cdot q}$ & &
      & \parbox[t]{10.7cm}{\raggedright Bjorken scaling variable.}\\
    $\boldsymbol{Q^2}$ & = & $-q^2$ & & 
      & \parbox[t]{10.7cm}{\raggedright Negative four-momentum 
      squared of the virtual photon.}\\
    $\boldsymbol{\nu}$ & = & $\frac{q\cdot P}{\sqrt{P^2}}$ 
                       & = & $E_e^{trf}-E_e^{trf\,\prime}$  
      & \parbox[t]{10.7cm}{\raggedright Energy of the virtual  
      photon in the target rest frame.}\\
    $\boldsymbol{y}$ & = & $\frac{q\cdot P}{k\cdot P}$ 
                     & = & $\frac{\nu}{E_e^{trf}}$ 
      & \parbox[t]{10.7cm}{\raggedright Fractional energy loss
      of the incident lepton.}\\
    $\boldsymbol{W^2}$ & = & $(P+q)^2$ & & 
      & \parbox[t]{10.7cm}{\raggedright  Invariant mass squared of 
      the hadronic final state.}\\
    $\boldsymbol{z_h}$   & = & $\frac{p_h\cdot P}{q\cdot P}$ & = & $\frac{E_h}{\nu}$ 
      & \parbox[t]{10.7cm}{\raggedright Fraction of the virtual photon 
      energy carried by the  hadron.}\\
    $\boldsymbol{p_T}$   & = & $|\vec{p}_{T}|$ & &
      & \parbox[t]{10.7cm}{\raggedright Hadron transverse momentum
        (relative to the virtual photon momentum).}\\\hline
  \end{tabular}\\
  \caption{Definitions of the kinematic variables for semi-inclusive
    DIS. The Lorentz invariant definition and its form in the target
    rest frame are provided. Particle 4-momenta are defined in
    Fig.~\ref{fig:NNDISkinematics}. All variables are
    experimentally measurable, hence typeset in boldface.
    Note that $x_B = Q^2 / (2M\nu)$
    independently of the chosen reference frame.}
  \label{tab:DISkinvar}
\end{table*}

\subsection{NN collisions} 
\label{sec:NNk}
In pQCD at leading order in the coupling constant $\alpha_s$, 
parton production in NN collisions proceeds through $2\ra
2$ partonic collisions (see Fig.~\ref{fig:NNDISkinematics} left and
Table~\ref{tab:DISkinvar} for the definition of kinematic variables.)
Several LO processes can contribute to a given $ij\ra f_1f_2$ collisions,
represented by a black disk in the cartoon, see
Ref~\cite{Field:1989uq} for details. The momenta of the 2 nucleons
colliding in the c.m.f. with energy $\sqrt s /2$ each are
\begin{align}
\begin{split}
  I & = \Big( \sqrt{\frac{\tilde s}{2}},\frac{M^2}{\sqrt {2\tilde s}}
    , \vec 0_T \Big) \\
  J & = \Big( \frac{M^2}{\sqrt{2\tilde s}}, \sqrt{\frac{\tilde s}{2}}
    , \vec 0_T \Big) 
  \label{eq:IJ}
\end{split}
\end{align}
where $M$ is the nucleon mass and
\begin{align}
  \tilde{s} = s \frac{1+\sqrt{1+M^4/s^2}}{2}  \ .
\end{align}
I will neglect terms of order $O(M^2/s)$ compared to terms of $O(1)$,
and will use $\tilde s \approx s$. Note also that in the 
definition of the nucleon momenta, I explicitly retain the nucleon
mass in Eq.~\eqref{eq:IJ} to be able to perform boosts to the rest
frame of either nucleon. 
If we assume the partons to be massless and collinear to their parent
nucleons, their 4-momenta in terms of the parton fractional momenta
$x_i$ read 
\begin{align}
\begin{split}
  i & = \Big( x_1 \frac{\sqrt s}{2}, 0, \vec 0_T \Big) \\
  j & = \Big( 0, x_2 \frac{\sqrt s}{2}, \vec 0_T \Big) \ .
\end{split}
\end{align}
In terms of rapidities and transverse
momentum $p_T$, the parton 4-momenta read
\begin{align}
  f_1 &= \Big(\frac{p_T}{\sqrt{2}}e^{y_1},\frac{p_T}{\sqrt{2}}e^{-y_1},-\vec{p}_T\Big) \\
  f_2 &= \Big(\frac{p_T}{\sqrt{2}}e^{y_2},\frac{p_T}{\sqrt{2}}e^{-y_2},\vec{p}_T\Big) \ .
\end{align}
We can express the parton fractional momenta in terms of $p_T,y_i$ as 
\begin{align}
\begin{split}
  x_1 & = \frac{p_T}{\sqrt{s}} (e^{y_1} + e^{y_2} ) \\
  x_2 & = \frac{p_T}{\sqrt{s}} (e^{-y_1} + e^{-y_2} ) \ .
\end{split}
\end{align}
Finally, the Mandelstam invariants are defined as follows,
\begin{align}
\begin{split}
  \hat{s} & = (i+j)^2 \\
  \hat{t} & = (i-f_1)^2 = (f_2-j)^2 \\
  \hat{u} & = (i-f_2)^2 = (f_1-j)^2
\end{split}
\end{align}
and 4-momentum conservation is expressed as
$
  \hat s + \hat t + \hat u = 0
$.
In terms of rapidities and transverse momentum, the Mandelstam
invariants read
\begin{align}
\begin{split}
  \hat{s} & = x_1 x_2 s \\
  \hat{t} & = -p_T^2 (1+e^{y_2-y_1}) \\
  \hat{u} & = -p_T^2 (1+e^{y_1-y_2}) \ .
\end{split}
\end{align}
In order to compare collider and fixed target experiments, and different 
beam energies, it is useful to consider the rapidity in the c.m.f.:
\begin{align}
  y_{c.m.f.} = y - y_{cm} \ .
\end{align}
The backward rapidity region (target hemisphere) 
corresponds to $y-y_{cm} < 0$, and the
forward rapidity region (projectile hemisphere) to $y-y_{cm}>0$.

Hadronization in the collinear factorization framework proceeds
through independent parton fragmentation into a hadron. It is
universal, i.e., independent of the process which produced the
fragmenting hadron, e.g., NN or DIS collisions
\cite{Collins:1981uk}. The hadron fractional momentum $z$ is defined
by
\begin{align}
  \begin{split}
    p_h^+ & = z f_1^+ \\
    \vec p_{hT} & = z \vec f_{1T} \ .
  \end{split}
\end{align}
Therefore the on-shell hadron momentum $p_h$ reads
\begin{align}
  p_h = (zf_1^+, \frac{m_h^2+z^2f_{1T}^2}{2 z f_1^+}, z\vec f_{1T}) \ .      
 \label{eq:zdef}
\end{align}
The parton and hadron rapidities are related by $y_1 = y_h +
\log(m_{hT}/p_{hT})$. The non perturbative dynamics of the
fragmentation process is encoded 
in universal fragmentation functions, which can be obtained in global
fits of experimental data \cite{Kniehl:2000fe,Kretzer:2000yf}.

\subsection{DIS collisions} 
\label{sec:DISk}
At LO in pQCD, deeply inelastic scattering proceeds by exchange of a
virtual photon in the $\hat t$-channel, explicitly shown in
Fig.~\ref{fig:NNDISkinematics} right. The DIS Lorentz invariants are defined in
Table~\ref{tab:DISkinvar}. 
Semi-inclusive nDIS is best discussed in terms of $\nu$ and $Q^2$,
which are the most relevant variables to hadron quenching processes in nuclear
targets. Analysis of inclusive DIS is usually carried
out using $x_B$ and $Q^2$.

DIS experiments can be performed with a fixed target (ft) or in collider
mode (cl). Examples are the EMC, HERMES, JLAB experiments, and the
Electron-Ion Collider (EIC), respectively. 
The colliding nucleon and lepton momenta are
\begin{align}\begin{split}    
  P_{ft} & = \Big( \frac{M}{\sqrt{2}}, \frac{M}{\sqrt{2}}, \vec 0_T \Big) 
  \ , \quad k_{ft} = \Big( \sqrt{2} E_e,0,\vec 0_T \Big) \\
  P_{cl} & = \Big( \frac{M}{2\sqrt{2}E_N},\sqrt{2} E_N, \vec 0_T \Big) 
  \ , \quad k_{cl} = \Big( \sqrt{2} E_e,0,\vec 0_T \Big) 
\end{split}\end{align}
where $E_e$ and $E_N$ are the electron and nucleon energies measured
in the laboratory frame. To discuss both modes at the same time, it is
convenient to introduce the target rest frame energy of the electron,
$\Etrf$:
\begin{align}
  \Etrf = 
    \bigg\{ \bay{ll}
      E_e & \text{fixed target} \\[.1cm]
      \frac{2E_NE_e}{M} \quad\ & \text{collider mode}
    \eay
\end{align}
Then the invariant $y$ for both modes becomes $y = \nu/\Etrf$.

\begin{figure*}[tb]
  \centering
  \includegraphics[width=8.5cm]{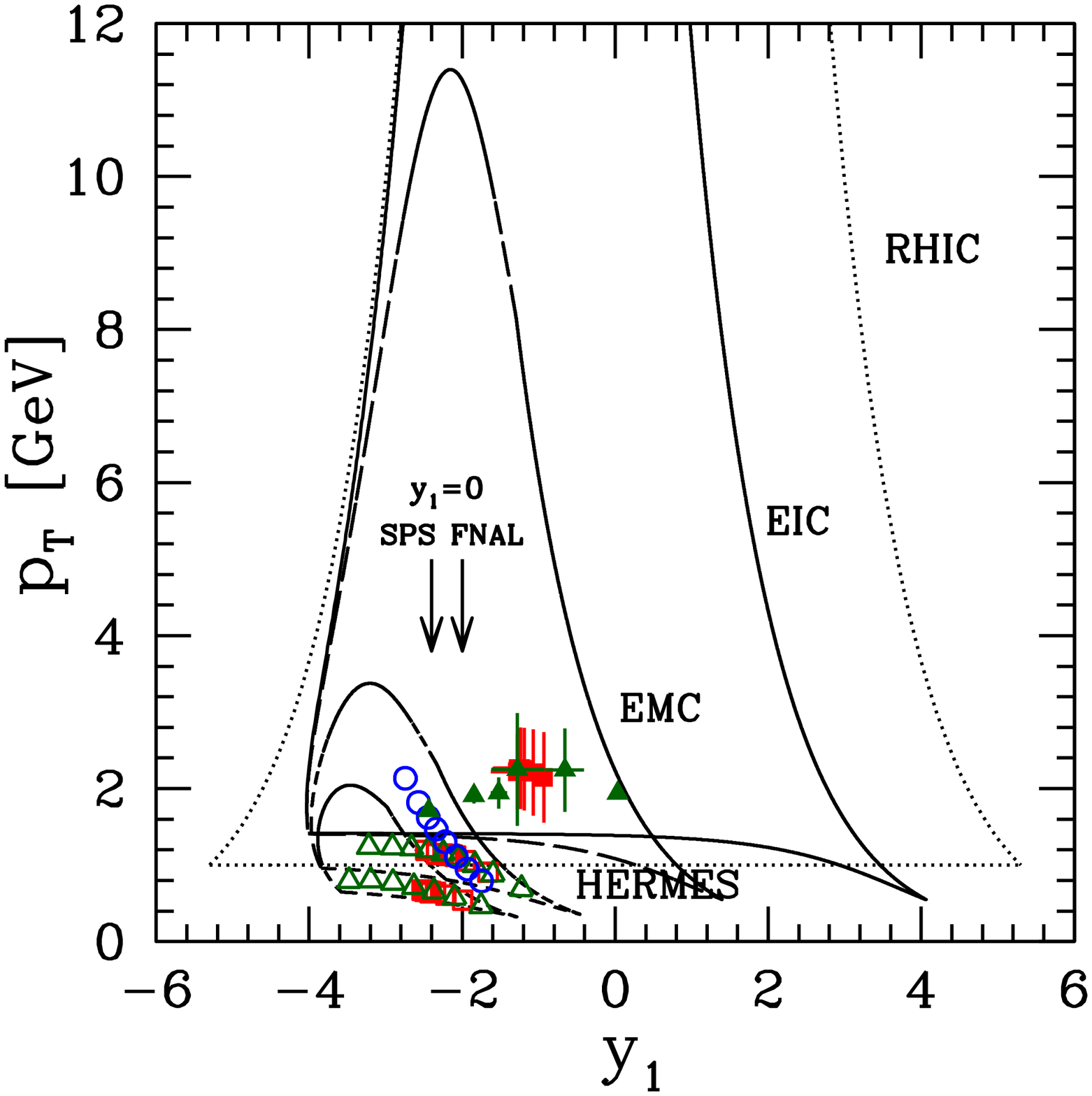}
  \includegraphics[width=8.5cm]{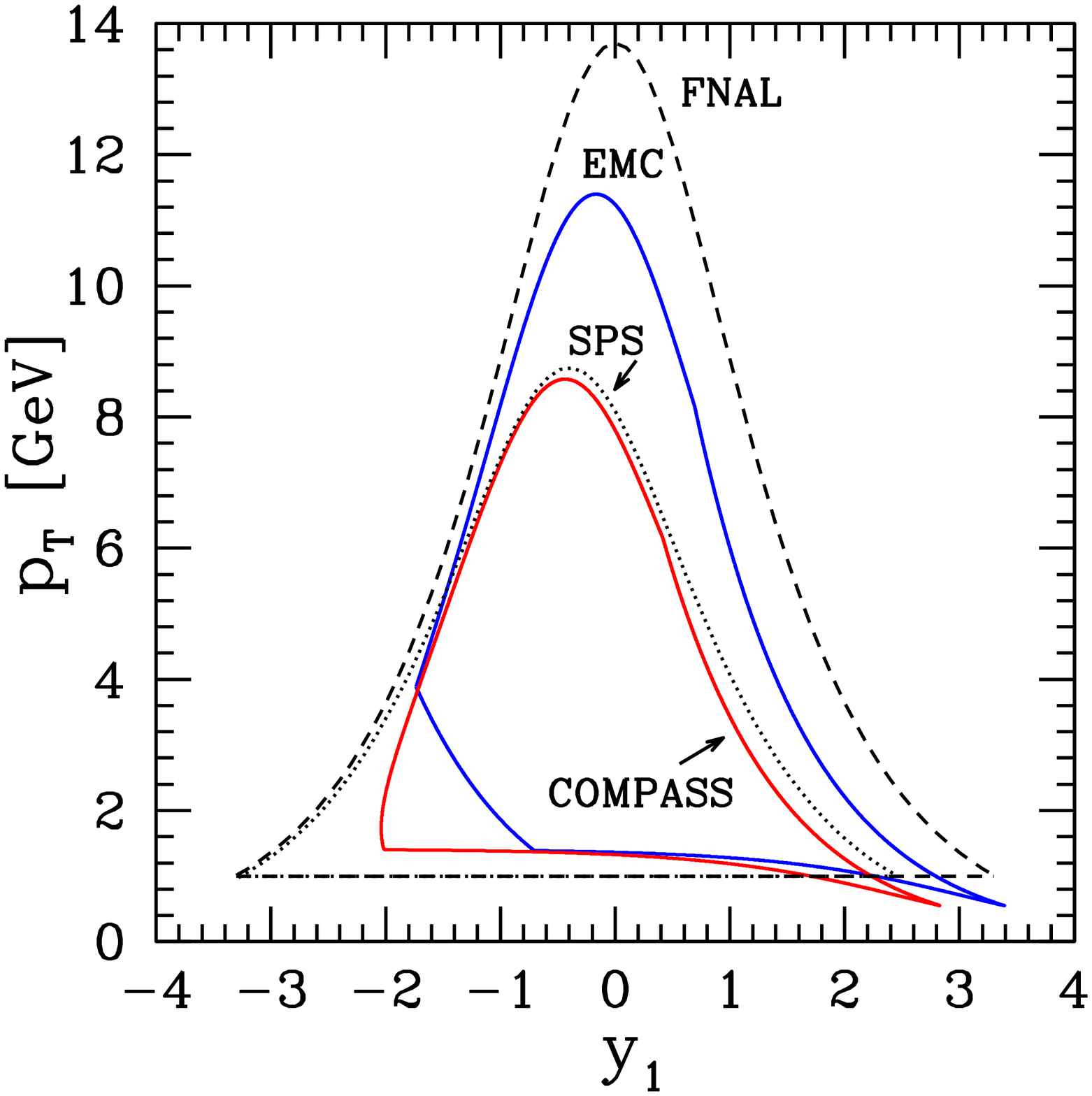}
  \caption{{\it Left:} the RHIC-equivalent phase space of DIS
    experiments at $E_e=27.6$ GeV (HERMES, solid line), at $E_e=12$
    GeV (HERMES and JLAB, dashed line), and at $E_e = 280$ GeV (EMC,
    dot-dashed line). The dotted line shows the borders of the LO pQCD
    phase space at top RHIC energy, $\sqrt{s}=200$ GeV. 
    The 2 arrows show the location 
    of the midrapidity region at SPS and FNAL fixed target
    experiments. The open and filled symbols 
    show the position of a representative
    selection  of data on hadron
    suppression collected at the HERMES
    \cite{Airapetian:2007vu,vanderNat:2003au} 
    and EMC experiments \cite{Ashman:1991cx}, respectively. 
    Red squares are for
    $z_h$-distributions, green triangles for $\nu$-distributions and
    blue circles for $Q^2$-distributions. The systematic uncertainties
    on EMC data, shown as error bars, are discussed in
    Section~\ref{sec:explessons}.  
    {\it Right:} NN-equivalent EMC and COMPASS phase space at
    $\sqrt{s}=27.4$ GeV, compared to the SPS and FNAL phase spaces. 
  }
  \label{fig:RHICequiv(HERMES)}
\end{figure*}

\subsection{The dictionary} 
\label{sec:dictionary}
To connect the DIS and NN kinematics, we can boost the DIS collision to a 
frame in which the target has energy $\sqrt{s}/2$ per nucleon.
Then, we can imagine the lepton to be a parton of a phantom nucleon of 
energy $\sqrt{s}/2$ collinear with the lepton, with 4-momentum 
$P'^\pm = P^\mp$. Comparing the top and bottom of
Fig.\ref{fig:NNDISkinematics} we can identify 
\begin{align}
\begin{split}
  P  & \equiv J , \quad  
  P' \equiv I , \quad  
  k  \equiv i , \quad  
  k' \equiv f_2 .
\end{split}
\end{align}
The virtual photon momentum $q$, the fractional momentum $x_e$ of the
initial state lepton  and the rapidity  $y_e$ of the final state lepton
are identified as follows 
\begin{align*}
\begin{split}
  q  & = k-k' \equiv i-f_2 , \quad
  x_e = k^+ / P'^+ \equiv x_1, \quad
  y_e \equiv y_2 \ .
\end{split}
\end{align*}
In this way, we can relate the DIS kinematics to the NN kinematics
discussed in Sect.~\ref{sec:NNk}. 
As an example, it is immediate to see that, in terms of NN variables,
$Q^2 = - \hat t$. The full translation dictionary from DIS to NN
variables can be obtained in a straightforward way 
by combining the results of
Sects.~\ref{sec:NNk}--\ref{sec:DISk} and the definitions of
Tables~\ref{tab:NNkvar}--\ref{tab:DISkinvar}. 

First, we can express the DIS invariants in terms of parton rapidities
and transverse momenta. Neglecting target mass corrections, i.e., up
to terms of $O(M^2/s)$, we obtain
\begin{align}
\begin{split}
  x_B & = \frac{p_T}{\sqrt{s}} (e^{-y_2}+e^{-y_1}) \\
  Q^2 & = p_T^2 (1+e^{y_1-y_2}) \\
  \nu & = \frac{p_T\sqrt{s}}{2M} e^{y_1} \\
  y   & = \frac{1}{1+e^{y_2-y_1}} \\ 
  z_h & = z \ . 
 \label{eq:DIS(NN)}
\end{split}
\end{align}
Note that the first 3 variables are not independent because 
$Q^2 = 2M x_B \nu$, and that $x_B=x_2$ is interpreted as the struck
parton fractional momentum, as expected in DIS at LO.
Note also that $\nu$ increases with increasing $p_T$ and increasing
$y_1$. In other words, a parton of positive and large $y_1$ travels in
the opposite direction as its parent nucleon, hence in the target rest
frame it is very fast. Conversely, a parton of negative and large $y_1$
travels in the same direction as its parent nucleon, which means quite
slow in the target rest frame. 
It is also interesting to note that up to terms of order $O(M^2/s)$,
the parton and hadron energy in the target rest frame are
\begin{align}
  E = \nu \qquad E_h = z_h \nu \ . 
\end{align}
Finally, we can invert
Eq.~\eqref{eq:DIS(NN)} to obtain the NN variables in 
terms of DIS invariants:
\begin{align}\begin{split}
  p_T^2 & = (1-y)Q^2 \\
  y_1 & = - \log \Big( \frac{Q\sqrt{s}}{2M\Etrf}
    \,\frac{(1-y)^{1/2}}{y} \Big) \\
  y_2 & = y_1 + \log \Big( \frac{1-y}{y} \Big) \\
  z & = z_h
  \label{eq:NN(DIS)1}
\end{split}\end{align}
with $y = \nu/\Etrf$. 

Note that in DIS, the electron energy $\Etrf$, hence the electron
$x_e$, is fixed by the
experimental conditions; this is different from NN collisions where
the parton $j$ has an unconstrained fractional momentum. 
Changing the c.m.f. energy to $\sqrt{s'}$ simply results in a shift of
the parton rapidity,
\begin{align}
  y_1 \xrightarrow[s\ra s']{} y_1 + \Delta y_1
  \label{eq:y1shift}
\end{align}
where
$
  \Delta y_1 = \log(\sqrt{s}/\sqrt{s'})
  \label{eq:Deltay1}
$.
The value of $\Delta y_1$ compared to RHIC top energy $\sqrt{s}=200$
GeV is listed in Table~\ref{tab:Deltay1} for
the experiments of interest in this paper.
Another difference between DIS and NN collisions is the rapidity
difference $\Delta y$ between the outgoing ``partons''.
In DIS, the electron fractional momentum is fixed, so that 
$
  \Delta y_{|DIS} = y_1-y_e = \log \big( y/(1-y) \big)
$
is determined for each $p_T$ and $y_1$ by the
corresponding value of $y=\nu/\Etrf$, and can span only a limited
range:
\begin{align}
  \log \Big( \frac{y_{min}}{1-y_{min}} \Big) \leq \Delta y_{|DIS} 
    \leq \log \Big( \frac{y_{max}}{1-y_{max}} \Big) \ .
\end{align}
For example, at HERMES the experimental acceptance $0.07 < y < 0.85$
translates into
$ 
  -1.1 < \Delta y_{|DIS} < 0.75 
$.
In NN collisions, neither parton fractional momentum is fixed by the
experimental conditions, hence $\Delta y_{NN}=y_1-y_2$ can span
\begin{align}
  -\log\Big( \frac{\sqrt{s}e^{-y_1}}{p_T} - 1 \Big) 
    \leq \Delta y_{|NN} \leq
  \log\Big( \frac{\sqrt{s}e^{y_1}}{p_T} - 1 \Big) \ .
\end{align}
For example, for an observed parton with
$y_1=-2$ and $p_T = 2$ GeV, corresponding to the middle of the HERMES
DIS phase space, we obtain
$
  -2.5 < \Delta y_{|NN} < 6.6 \ ,
$
even though the average $\vev{\Delta y_{|NN}}$ will lay 
in the middle of this interval.

\section{Comparing the phase spaces}
\label{sec:phasespaces}
We can now compare in detail the phase spaces for parton production in
NN and DIS collisions. For this purpose, I will define a 
NN-equivalent DIS phase space and a DIS-equivalent NN phase space.

\begin{table}[tb]
  \centering
  \begin{tabular}{c|ccccc}
                     & SPS  & FNAL & RHIC & RHIC & LHC   \\\hline
    $\sqrt{s}$ [GeV] & 17.5 & 27.4 & 63   & 200  & 5500  \\
    $\Delta y_1$     & 2.4  &  2.0 &  1.2 & 0    & -3.3   \\
  \end{tabular}
  \caption{
    Rapidity shifts
    $\Delta y_1$ of the RHIC-equivalent DIS phase space, tabulated for
    some energies of interest.}
  \label{tab:Deltay1}
\end{table}

\subsection{NN-equivalent DIS phase space}
\label{sec:NNequiv}
Given a DIS phase space, i.e., a given experiment acceptance
region in the $(\nu,Q^2)$ plane, I define its
{\it NN-equivalent phase space} as its image
in the $(p_T,y_1)$ under Eqs.~\eqref{eq:NN(DIS)1}. (I do not consider
the transformation of the fragmentation variable $z_h$ into $z$
because of its triviality.) 
The reason for this definition is that for both NN and DIS
collisions we can identify the parton $f_1$ of
Fig.~\ref{fig:NNDISkinematics} with the ``observed'' parton in NN and
DIS collisions, i.e., the parton which fragments into the observed 
hadron. Then the variables $p_T$ and $y_1$ fully 
characterize the observed parton. An analogous
definition holds when using $x_B$ instead of $\nu$ as independent
variable. 

As an example, the HERMES DIS phase space in the
$(\nu,Q^2)$ plane is
determined by the values of $W^2_{min}$, $Q^2_{min}$ and $y_{max}$:
\begin{align}\begin{split}
  & \frac{Q^2_{min}+W^2_{min}-M^2}{2M} \leq \nu \leq y_{max}\,\Etrf \\
  & Q^2_{min} \leq Q^2 \leq M^2+2M\nu - W^2_{min} \ .
    \label{eq:DISps} 
\end{split}\end{align}
Additionally, one may impose stronger cuts on $\nu$, e.g., $\nu\geq
\nu_{min}$, as at the EMC experiment, and in some HERMES analysis.
 
With Eqs.~\eqref{eq:NN(DIS)1} it is easy to plot the NN-equivalent
DIS phase space in the $(y_1,p_T)$ plane. 
As an example, we can consider the RHIC-equivalent phase space of the
HERMES and EMC experiments, using $\sqrt{s}=200$ GeV, shown
in Fig.~\ref{fig:RHICequiv(HERMES)} left.
Note that according to Eq.~\eqref{eq:y1shift}, 
the NN-equivalent phase space at other center of mass energies 
can be obtained by a shift $y_1 \ra y_1+\Delta y_1$, see
Table~\ref{tab:Deltay1}.  
I assume the pQCD formulae used to define the NN-equivalent phase
space to be valid at RHIC top energy for 
$p>p_0=1$ GeV: the corresponding pQCD
confidence region is plotted as a dotted line, see
Eq.~\eqref{eq:NNps} for details.

We can see that the HERMES experiment, with $\Etrf = 12$ and 27.6 GeV,
covers less than one third of the available RHIC $p_T$ range 
at $y_1 \approx -3$, with shrinking $p_T$ coverage at
larger rapidity. In the SPS/FNAL midrapidity region it reaches $p_T =
2.5$ GeV at most.
Since 
\begin{align}\begin{split}
  y_1 & \leq \log \Big( \frac{\sqrt{s}}{2M\Etrf} 
    \frac{p_T}{y_{max}}\Big)
\end{split}\end{align}
and $y_{max}$ cannot be increased above 1, 
the only way to effectively reach larger values of $y_1$
is to increase the electron beam energy $\Etrf$. Indeed, 
the EMC experiment, with $\Etrf = 100-280$ GeV, covers a larger span
in rapidity and extends to $y_1 \gtrsim 0$.  
Moreover, the increased energy allows in principle to reach much
higher $p_T$ than at HERMES. However, only the $p_T \lesssim 3$ GeV
region has been explored. 
As also shown in Fig.~\ref{fig:RHICequiv(HERMES)} left, the
proposed Electron-Ion Collider (EIC)
\cite{Deshpande:2005wd,EICeAwhite} 
will be able to effectively study the $y_1 > 0$ region, and cover most
of the RHIC phase space. Likewise, it will cover only the $y_1<0$
part of the LHC phase space.

The reason why present experimental data in $\ell+A$ collisions reach
only $p_T \lesssim 2$ GeV, is that conventional DIS variables $z_h$,
$\nu$ or $Q^2$ explore the available NN-equivalent phase space in an
uneven way. Moreover, in single differential distributions like
$dN^h_A/dz$, the integration over the remaining variables favors low
values of $Q^2$, hence low-$p_T$ values. While HERMES is inherently
limited in its $p_T$ reach by the low electron beam energy, the EMC
experiment covers, in principle, most of the SPS and FNAL phase
space, see Fig.~\ref{fig:RHICequiv(HERMES)} right. Therefore,  
a rebinning of the EMC experimental data in terms
of NN variables would result in an experimental measurements of final
state nuclear effects, much needed for correctly interpreting large-$p_T$
hadron spectra in $h+A$ and $A+A$ collisions at SPS and FNAL.
Another possibility would be to study $\mu+A$ collisions at the
COMPASS experiment \cite{Abbon:2007pq}, 
which has a muon beam energy of $E_\mu=160$ GeV
comparable to EMC, and whose phase space is also shown in the plot.

\begin{figure*}[tb]
  \centering
  \includegraphics[width=6.5cm]{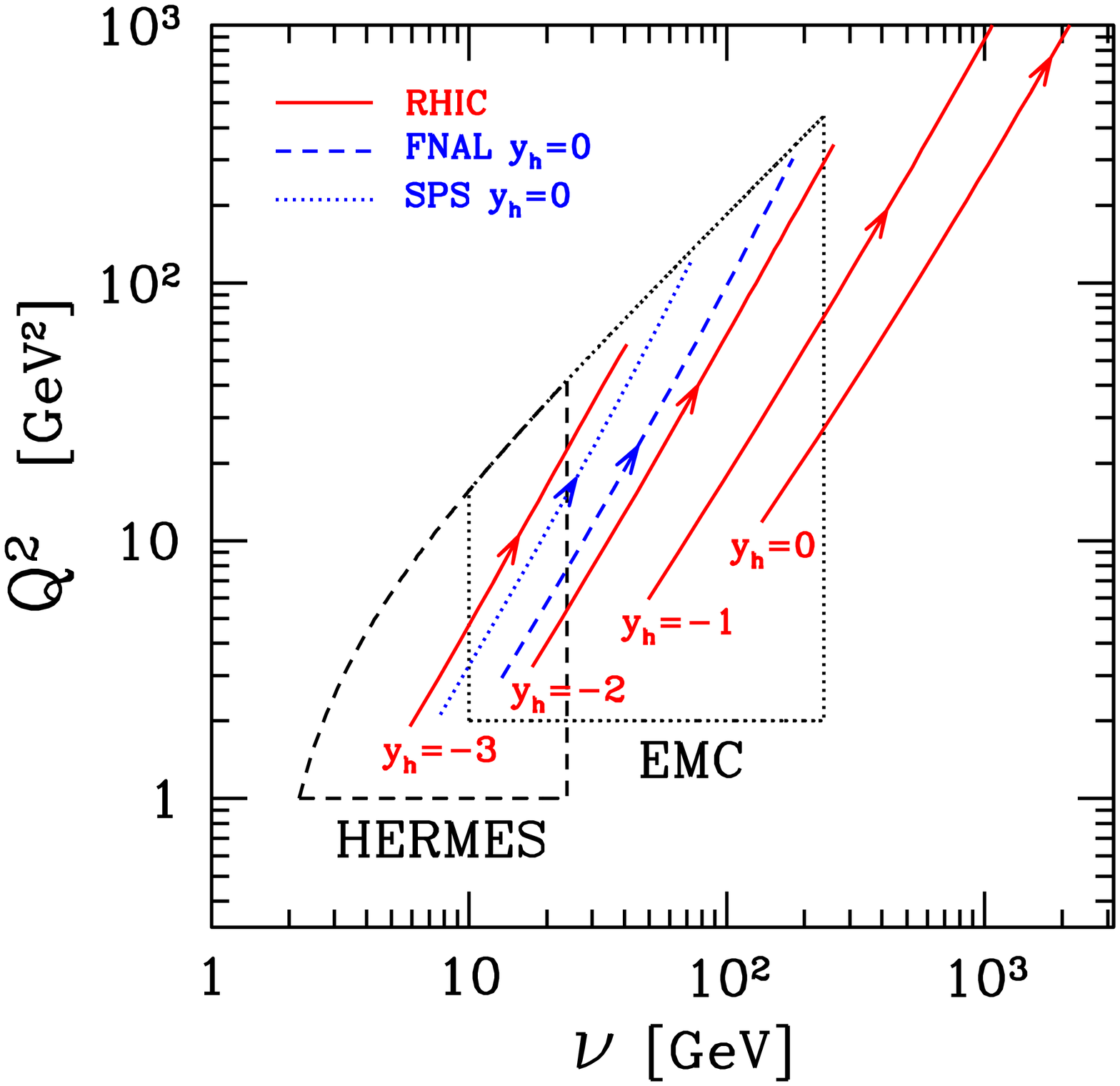}
  \includegraphics[width=6.5cm]{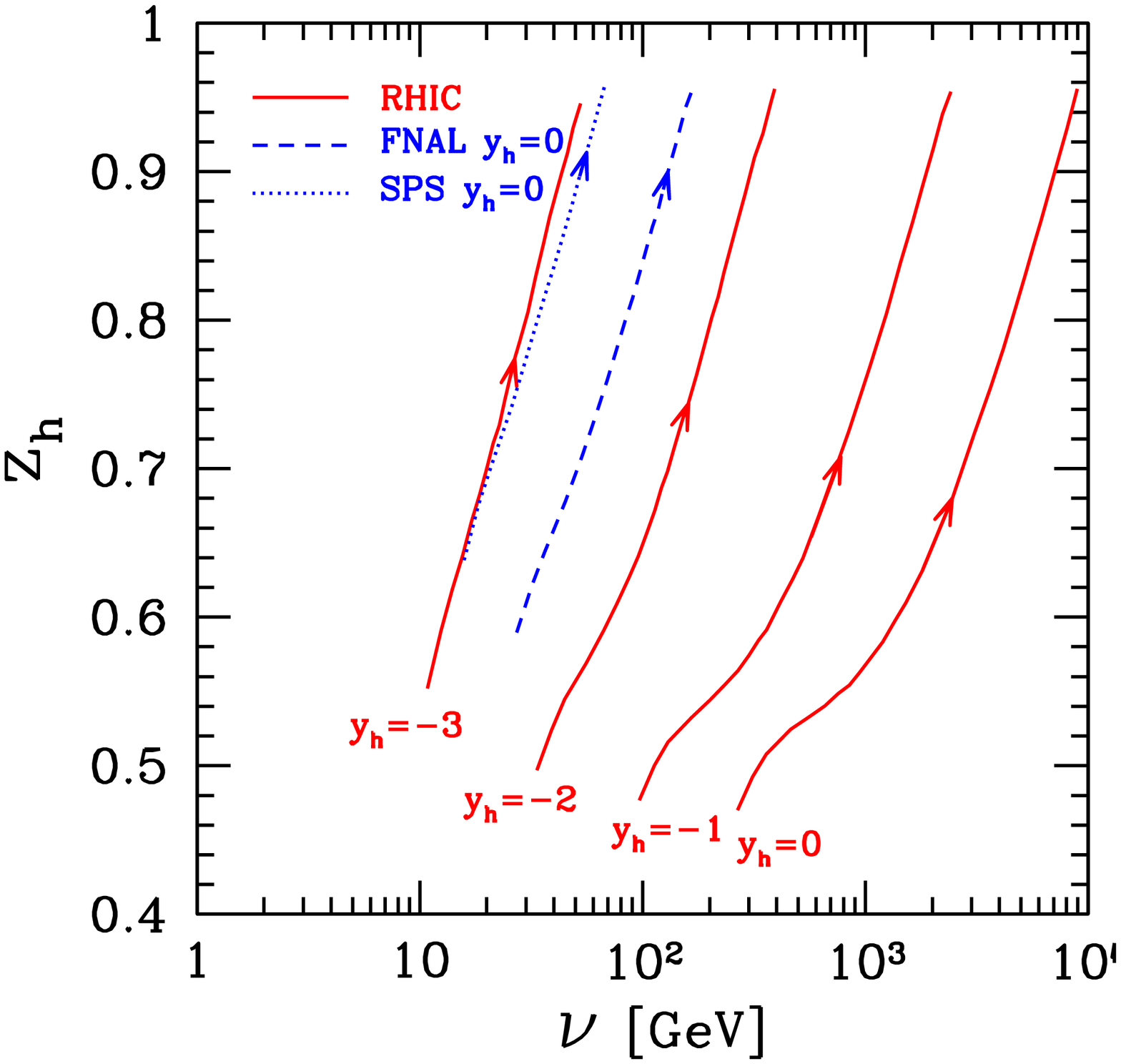}
  \caption{{\it Right:} Fixed-$y_1$ NN trajectories 
    plotted in the DIS-equivalent $(\nu,Q^2)$ 
    phase space for RHIC at $\sqrt{s}=200$ GeV and various rapidities,
    for FNAL and SPS at midrapidity. The dot-dashed line encloses the
    HERMES phase space; the dashed line encloses the EMC phase
    space. The arrow indicates the direction of increasing $\vev{p_T}$
    and $\vev{z_h}$. {\it Left:} Trajectories in the $(\nu,z_h)$
    plane. The arrows indicate increasing $p_T$ and $Q_2$.}
  \label{fig:HERMES(RHIC)}
\end{figure*}

\subsection{DIS-equivalent NN phase space}
\label{sec:DIS-eqNN}
When discussing NN collisions in the framework of collinear
factorization in pQCD, we should first define the region of validity
of perturbative computations: $p_T \geq p_0$. Typically 
one needs $p_0 \gtrsim 1\text{\ GeV}$, which agrees with
the phenomenological analysis of
Refs.~\cite{Eskola:2002kv,Accardi:2003jh}. 
Then, the NN phase space at a given
$y_1$ is defined by the kinematic bounds on $2\ra 2$ parton
scatterings \cite{Eskola:2002kv}:
\begin{align}\begin{split}
  & |y_1| \leq \cosh^{-1} \Big(\frac{\sqrt s}{2p_0}\Big) \\
  & p_0 \leq p_T \leq \frac{\sqrt s}{2 \cosh(y_1)} \\
  & -\log \Big( \frac{\sqrt{s}}{p_T}-e^{-y_1} \Big) 
    \leq y_2 \leq 
    \log\Big( \frac{\sqrt{s}}{p_T}-e^{y_1} \Big) \\
  & \frac{m_{hT}}{\sqrt{s}} e^{y_h} 
    \Big(1+\frac{p_{hT}^2}{m_{hT}^2e^{y_h}} \Big)   \leq z \leq 1
  \label{eq:NNps}
\end{split}\end{align}
Introduction of intrinsic parton transverse momentum in the formalism,
or use of next-to-leading order kinematics \cite{Guzey:2004zp},
would relax somewhat these bounds. We should also keep in mind that at
large rapidity, where the $2\ra2$ phase space is becoming more and
more restricted, $2\ra 1$ parton fusion processes may become the
dominant mechanism, because they are sensitive to much
lower fractional momenta $x_i$ \cite{Accardi:2004fi}. Hence, at the
boundary of the NN phase space, the presented analysis becomes
unreliable. 

The {\it DIS-equivalent NN phase space} is defined as the image of
Eqs.~\eqref{eq:NNps} in the $(\nu,Q^2,y,z_h)$ space under
Eqs.~\eqref{eq:DIS(NN)}. It is 4-dimensional and
difficult to directly visualize. A way around this problem is to
define suitable trajectories in NN phase space averaged over $y_2$, 
and to project them into the DIS-equivalent ($\nu$,$Q^2$) and
($\nu$,$z_h$) phase spaces. We can define a
$p_{hT}$- and $y_h$-dependent average observable as follows 
\begin{align}
  \vev{\OO}_{p_{hT},y_h} = 
    \frac{ 
      \int dz\,dy_1\,dy_2\, \OO(p_T,y_1,y_2,z) 
      \frac{d\hat\sigma^{AB\ra hX}}{dp_T^2dy_1dy_2dz}
    }{
      \int dz \,dy_1\,dy_2
      \frac{d\hat\sigma^{AB\ra hX}}{dp_T^2dy_1dy_2d_z}
    } \ ,
\end{align}
where
\begin{align}
  \frac{d\hat\sigma^{AB\ra hX}}{dp_T^2dy_1dy_2dz} 
    = \sum_{f_1} \frac{1}{z^2} D_{f_1}^h(z) 
    \frac{d\hat\sigma^{AB\ra f_1X}}{dp_T^2dy_1dy_2} \ ,
\end{align}
$d\hat\sigma^{AB\ra f_1X}$ is the LO pQCD differential cross-section
for production of a $f_1$ parton pair in a collision of hadrons
$A$ and $B$ (nucleons or nuclei), and $D_{f_1}^h$ is its
fragmentation function into the observed hadron,
see Ref.~\cite{Eskola:2002kv} for
details \footnote{The computations presented in this paper, differ
  from \cite{{Eskola:2002kv}} 
  in 2 respects: I defined $z=p_h^+/f_1^+$ instead of $z=E_h/E_{f_1}$,
  and I regularized the pQCD cross-section in the infrared with a shift
  $p_T^2\ra p_T^2+p_0^2$ instead of using a sharp cutoff $p_T>p_0$. The
  difference is mostly seen at small $p_{hT} \lesssim p_0$.}.
Then, we can use Eqs.~\eqref{eq:DIS(NN)} to compute
$\vev{\nu}_{p_{hT},y_h}$, $\vev{Q^2}_{p_{hT},y_h}$, and
$\vev{z_h}_{p_{hT},y_h}$. These values parametrize the fixed-$y_h$
trajectories 
$
  \{ (\vev{\nu}_{p_T,\bar y},\vev{Q^2}_{p_T,\bar y})
    ; p_T\geq p_0 \}  
$ 
and 
$
  \{ (\vev{\nu}_{p_T,\bar y},\vev{z_h}_{p_T,\bar y})
    ; p_T\geq p_0 \}  
$
in the DIS-equivalent phase space.

As an example, in Fig.~\ref{fig:HERMES(RHIC)} 
I considered NN collisions at RHIC top energy $\sqrt{s}=200$ GeV 
and at fixed target energies $\sqrt{s}=17-27$ GeV, and plotted the
fixed-$y_h$ trajectories in the DIS-equivalent phase space.
The range of $p_T$ spanned along each
trajectory is tabulated in Table~\ref{tab:pTzhrange}.
The spanned range in $Q^2$ is limited by the maximum $p_T$ at 
each rapidity, according to Eq.~\eqref{eq:NNps}. 
As expected, the larger the rapidity $y_h\approx y_1$ 
the smaller the spanned $\nu$. RHIC trajectories with 
$y_h\lesssim -2$ span pretty low values of $\nu \lesssim 60$ and large
values of $z_h \gtrsim 0.5$, where the EMC
and HERMES experiments have shown non negligible cold nuclear matter
suppression of hadron production. At higher rapidity, the larger spanned 
values of $\nu$ will make cold nuclear matter effects less prominent.
The consequences of these remarks for the interpretation
of hadron production in $h+A$ and $A+A$ collisions will be further
discussed in Section~\ref{sec:coldjetquenching} and \ref{sec:ISvsFS}. 

\begin{table}[tb]
  \centering
  \begin{tabular}{c|c|c|cccc|}
     & SPS 
     & FNAL 
     & \multicolumn{4}{c|}{RHIC} \\
     & $\sqrt{s}=17.5$ GeV
     & $\sqrt{s}=27.4$ GeV
     & \multicolumn{4}{c|}{$\sqrt{s}=200$ GeV} \\\hline  
 $y_h$          & 0     & 0     & 0     & -1    & -2    & -3   \\
 $p_{hT}$       & 1--8  & 1--12 & 1--90 & 1--60 & 1--25 & 1--9  \\\hline  
  \end{tabular}
  \caption{Range of average $\vev{pT}$ spanned along RHIC
    trajectories at fixed rapidity $y_1$ and $\sqrt{s}=200$
    GeV. $p_{hT}$ is quoted in GeV.}
  \label{tab:pTzhrange}
\end{table}

\section{Final state cold nuclear quenching in $\boldsymbol{h+A}$
  collisions.} 
\label{sec:coldjetquenching}

As we have seen, a parton produced at negative rapidity, $y-y_{cm}<0$,  
in a $h+A$ collision travels in the same direction as the
target nucleus: seen in the nucleus rest frame, it appears to move
slowly and corresponds to a low value of $\nu$ in the language of
$\ell+A$ collisions. Therefore, 
based on the observed suppression of hadron production in
lepton-nucleus DIS
\cite{Airapetian:2003mi,Airapetian:2000ks,Airapetian:2003mi,Ashman:1991cx,
Osborne:1978ai}   
at low $\nu$, and on the kinematic analogy between DIS and NN
collisions discussed in the previous sections, we can expect
non-negligible hadron suppression due to FS interactions in cold
nuclear matter also in $h+A$ and $A+A$ collisions.

\begin{figure}[bt]
  \centering
  \includegraphics[width=0.9\linewidth]{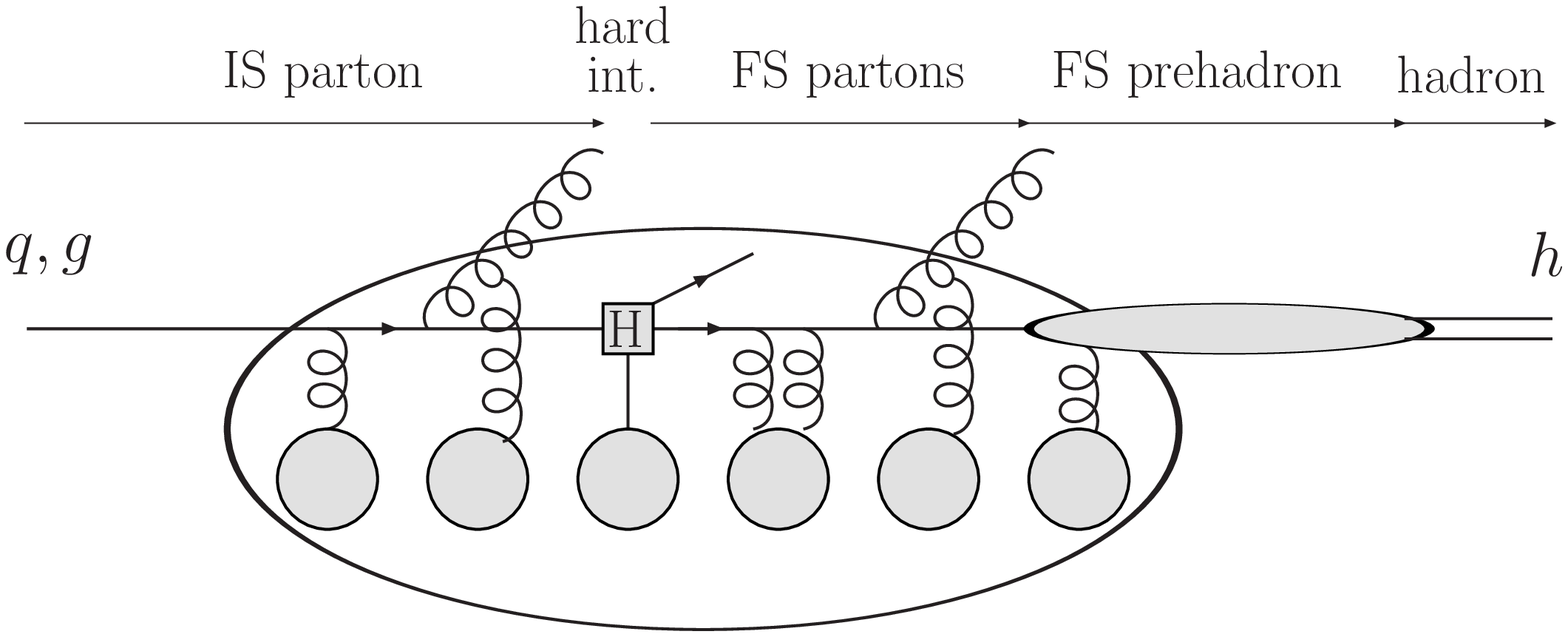}
  \vskip.5cm
  \includegraphics[width=0.9\linewidth]{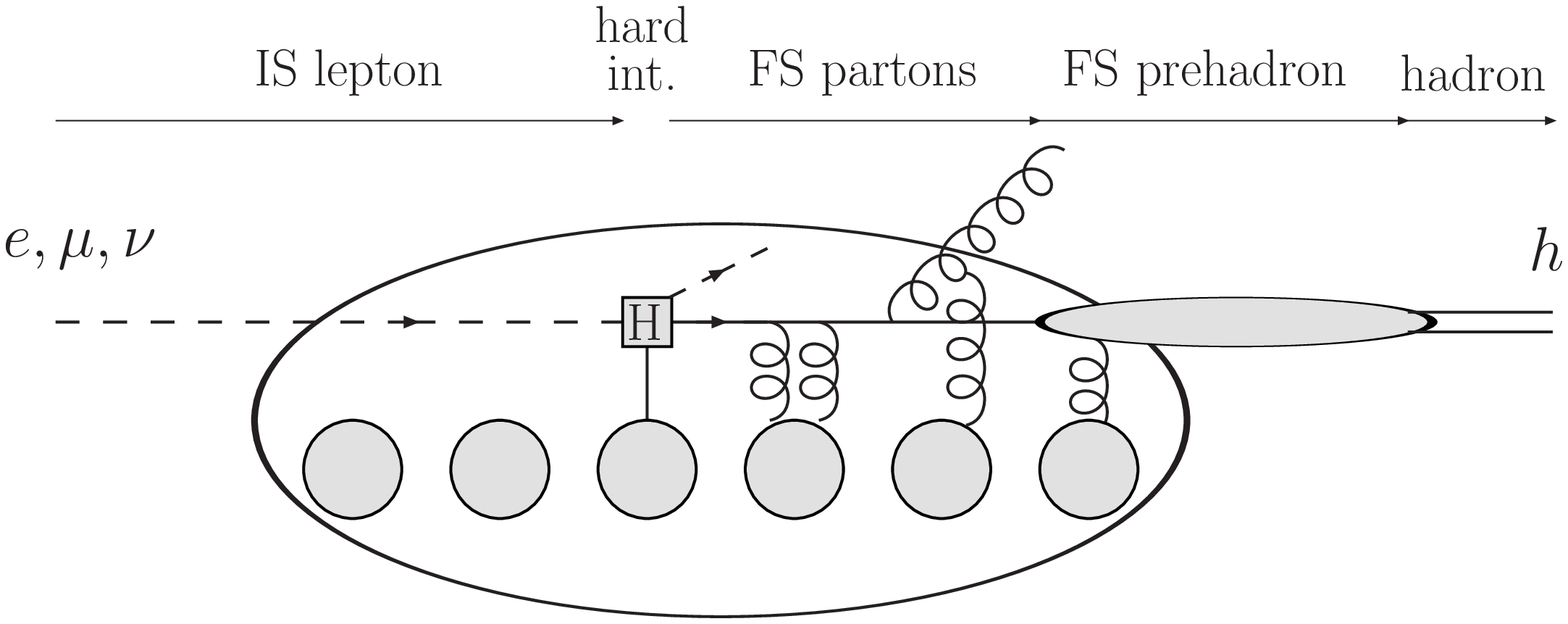}
  \caption{Top: Initial and final state interactions in $h+A$ collisions in
    the nucleus rest frame. Bottom: absence of initial state
    interactions in $\ell+A$ collisions. 
    The nucleus is drawn as an oblong oval for
    convenience only.}
  \label{fig:ISFS}
\end{figure}

Discussion of medium effects is best carried out in the medium rest
frame: in the case of cold nuclear matter in $\ell+A$ and $h+A$
collisions it is the nucleus rest 
frame. I am interested here in processes characterized by large
values of $x_B \equiv x_2 \gtrsim 0.1$, typical of semi-inclusive nDIS
measurements at HERMES and large $p_T$ hadron production at not too
forward rapidity in $h+A$ collisions. In this regime, the hard
interaction is well localized inside the nucleus, and the nucleons act
incoherently as targets \cite{Hoyer:1995gd}. The process evolves in
time as follows, see Fig.~\ref{fig:ISFS}. First
the electron (or a parton belonging to the proton) 
penetrates the nucleus, and undergoes a localized hard
collisions. Then, a ``final-state'' system  of 1 electron and 1
parton (or 2 partons) is produced, with both particles essentially
traveling along the beam direction, even for rapidity values far from
the center of mass rapidity in the target hemisphere ($y-y_{cm} < 0$). 
Later on the final state partons hadronize and one of the produced
hadrons is detected. 
The time scale on which hadronization starts
after the hard interaction is not at present well known
\cite{Accardi:2006ea}; it may be as small as the nuclear radius
\cite{Accardi:2006qs,Kopeliovich:2003py,Airapetian:2007vu}, in which
case the hadronization process would start in the medium.
Nuclear medium effects may be classified as initial state (IS) effects
on particles before the hard interactions, or final state (FS) effects
on particles created after the hard interaction.
In the case of $\ell+A$ collisions, electromagnetic reinteractions of the
incoming or outgoing lepton are suppressed compared to the strong
FS reinteraction of the parton and hadronizing system. For $h+A$
collisions one needs in principle 
to account for both IS and FS interactions \cite{Vitev:2007ve}.

\begin{figure*}[tb]
  \centering
  \includegraphics[width=6.5cm]{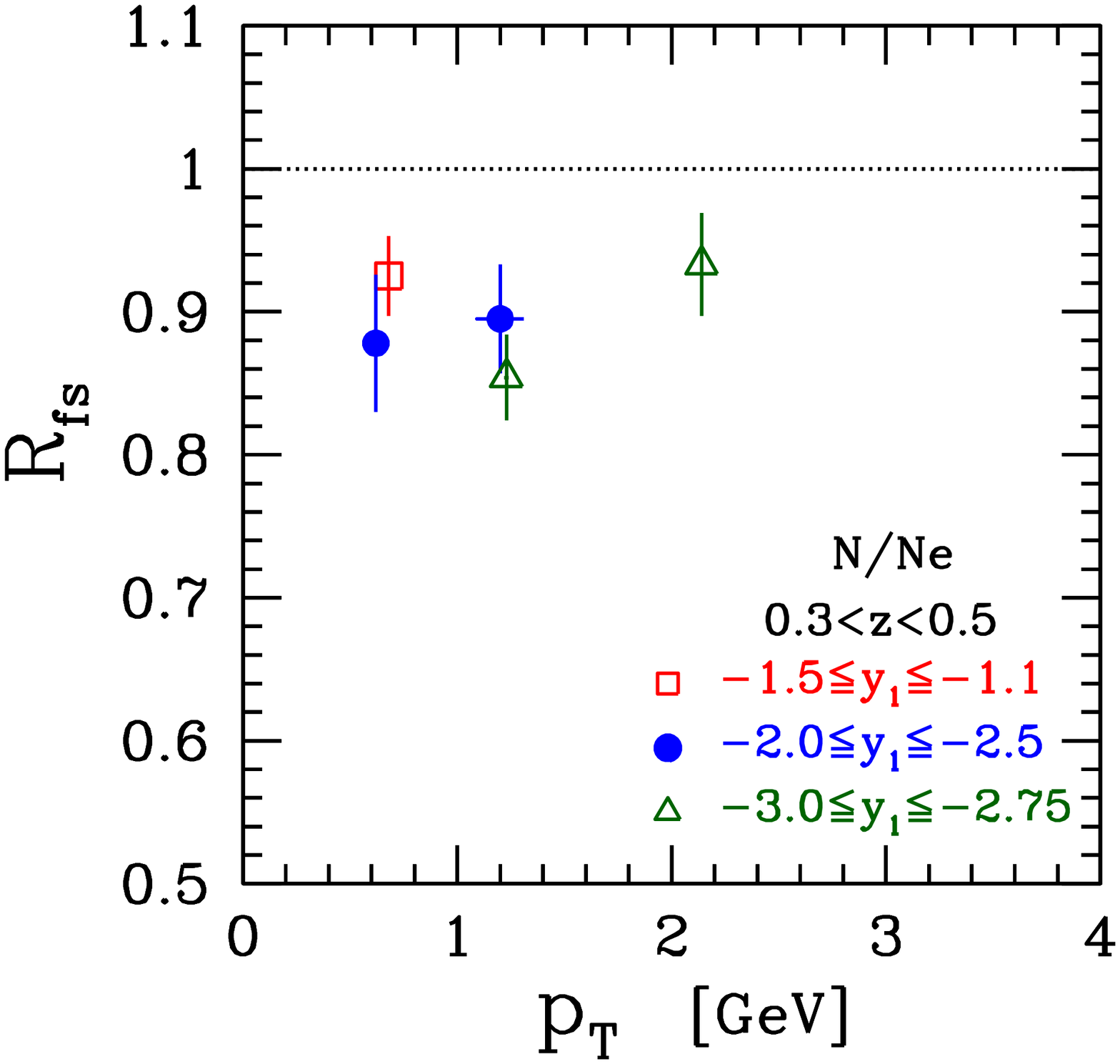}
  \includegraphics[width=6.5cm]{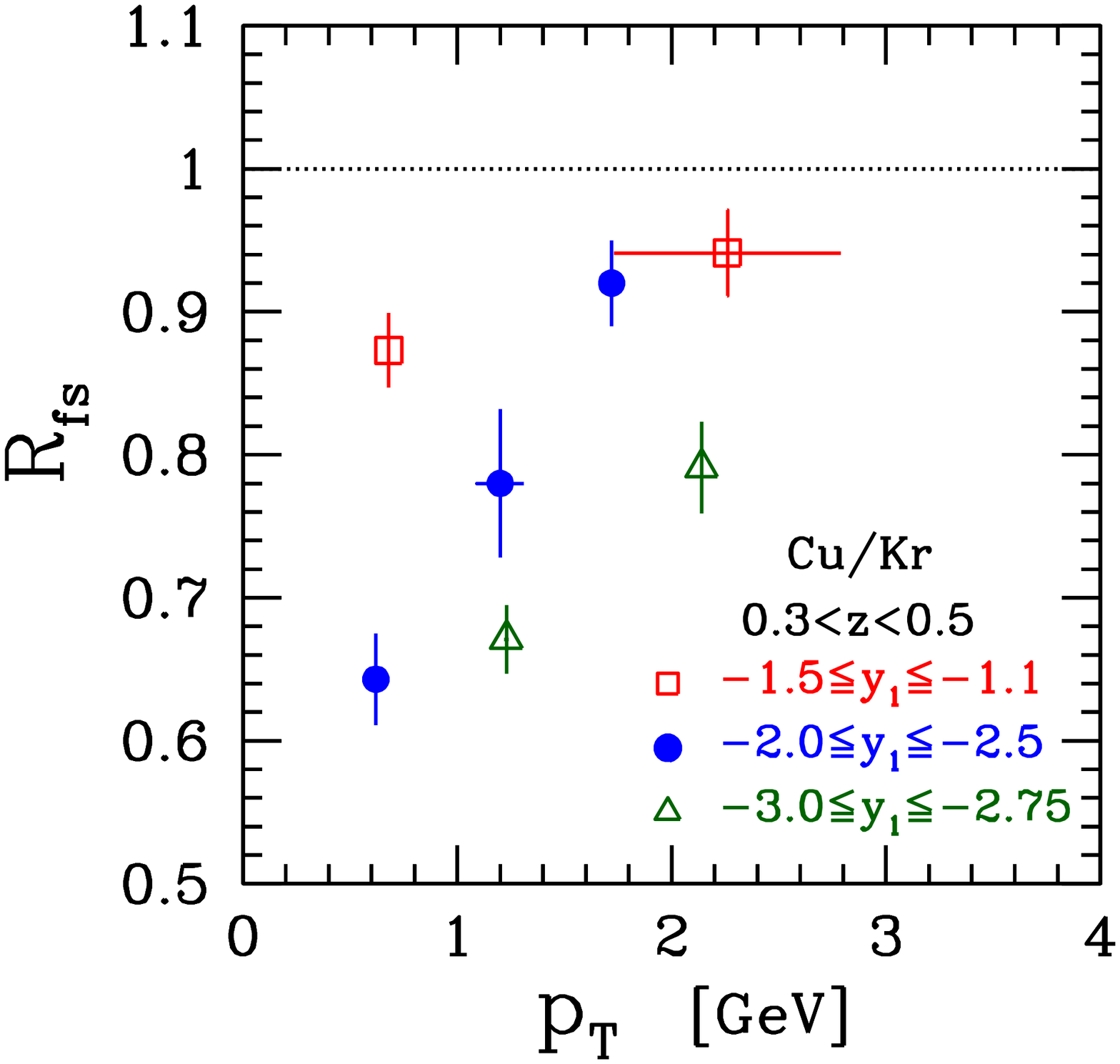}
  \caption{Cold jet quenching in d+A collisions on light and heavy
    targets at $y_1\approx -1.3, -2.25, -3.0$ and $z\approx 0.4$, 
    obtained from HERMES and
    EMC data on heavy and light targets. See main text for details.}  
  \label{fig:coldquenchexp}
\end{figure*}

Except at very forward rapidity, 
I will assume FS and IS effects in $h+A$ collisions to
be factorizable because of the large rapidity difference between the
IS and FS partons induced by the hard scattering. Then, I 
will explore the
possible size of FS effects on single inclusive hadron production.
Differently from $\ell+A$ collisions, the proton projectile interacts
with the nucleons along its trajectory. The hard parton produced in
the hard collision starts propagating at nearly the speed of light in
the same direction but slightly behind the projectile proton. 
The time scale for parton production,
$t_{hard}\propto 1/Q$ is much smaller than the time scale for soft
particle production in proton-nucleon collisions, $t_{soft}\propto
1/\Lambda_{QCD}$. Hence, we may assume the nuclear medium traversed by
the produced parton in $h+A$ collisions to have approximately 
the same properties as 
the cold nuclear matter which would be traversed in $\ell+A$ collisions,
i.e., the target nucleus itself. With this in mind, we may assume
final state hadron quenching effects to be comparable in the 2
cases.

\subsection{Lessons from $\boldsymbol{\ell+A}$ data}
\label{sec:explessons}

Nuclear modifications  of hadron production in $\ell+A$ collisions are
typically studied in terms of the hadron multiplicity ratio
\begin{align}
  R_M^h(z_h,nu,Q^2) = \frac{1}{N_A^{DIS}}\frac{dN_A^h}{dz_h d\nu dQ^2} \Bigg{/}
    \frac{1}{N_D^{DIS}}\frac{dN_D^h}{dz_h d\nu dQ^2}  ,\
    \label{MultiplicityRatio}	   
\end{align}
i.e., the single hadron multiplicity on a target of mass number $A$ 
normalized to the multiplicity on a deuteron target. 
Then, we can use the dictionary \eqref{eq:NN(DIS)1} 
and plot $R_M^h$ measured in $\ell+A$ collisions
as a function of the kinematic variables $p_T$, $y_1$ and $z$. This
will give a rough estimate of final state effects in $h+A$ collisions.
The results are presented in Fig.~\ref{fig:coldquenchexp}, and the
procedure used is discussed below. 

Data on $R_M$ are usually binned
in either $z_h$, $\nu$ or $Q^2$. Except for the EMC data, they are
presented alongside the average value of the unbinned variables. 
For HERMES data \cite{Airapetian:2000ks,Airapetian:2003mi,
Airapetian:2007vu}, I used the experimentally 
measured values of the DIS variables 
to compute the equivalent $p_T$, $y_1$ and $z$. For EMC data
\cite{Ashman:1991cx}, I used a
computation of the average unbinned variables from the GiBUU Monte
Carlo generator \cite{Gallmeister:2007an,Gallmeister}, which was shown
to well reproduce the corresponding measurements at HERMES
\cite{Falter:2004uc}. Another complication
arises from the fact that EMC data have been obtained by averaging
measurements at 3 electron beam energies, $E_e^{trf} = 100$ GeV, 200
GeV and 280 GeV; however, the details of such averaging are not
immediately clear from the original paper. Therefore, I used the minimum and
maximum of the computed average variables to obtain the corresponding
minimum and maximum of the NN variables, considered as error band
around their average value. Data have been selected to fall into 3
bins in $y_1$ ($-3.0 \leq y_1 \leq -2.75$, $-2.5 \leq y_1 \leq -2.0$, 
and $-1.5 \leq y_1 \leq -1.1$), and 1 bin in $z$ ($0.3\leq z \leq 0.5$).
The choice of $y_1$ bins has been made in order to minimize the spread of
$y_1$ and $z$ inside the bin, and to keep it as much as possible balanced
around the middle value. The chosen $z$ bin is the richest in measured
data. Furthermore, data with similar $p_T$ and from the same
target have been combined, with an error band in both $p_T$ and $R_M$ 
corresponding to the highest data plus error value and lowest data
minus error value, the central value being placed in the middle.
Of course, this procedure is a poor man's substitute for direct
experimental  binning in $p_T$ and $y_1$. 

The results of Fig.~\ref{fig:coldquenchexp} clearly show the evolution
of final state cold nuclear quenching with rapidity: the quenching increases
with decreasing $y_1$. This was expected from the kinematic analysis
of Section~\ref{sec:LO}, which shows a decreasing $\nu$ with
decreasing rapidity. The size of hadron quenching is not small,
especially for large nuclei and small $y_1$ rapidity. 
Its evolution with $z$ is not 
shown in the figure because of large overlapping error bars arising in
the rebinning procedure. However, the original HERMES and EMC
$z_h$-distributions clearly show an increasing quenching with
increasing $z_h$, especially at large $z_h \gtrsim 0.5$, where most of
hadron production in $h+A$ collisions takes place. Note also that
quenching increases with the target atomic number. 

As evident from Fig.~\ref{fig:coldquenchexp}, the $p_T$ range covered
by HERMES and EMC is quite limited compared to the $p_T$ for which
hadron production in $h+A$ and $A+A$ can be measured. As remarked in
Section~\ref{sec:NNequiv} this situation can be improved with a
rebinning of EMC data, or with new measurements of hadron attenuation
in $\mu+A$ collisions at the COMPASS experiment, which can in
principle reach up to $p_T \approx 8-10$ GeV. 

\subsection{Theoretical estimate for $\boldsymbol{h+A}$ collisions}

As already remarked, in DIS, one has experimental control over all the
kinematic variables. In h+A collisions $Q^2$ and, most
importantly for our considerations, $z$ are not experimentally
accessible. The non-trivial correlation of these variables with the
measurable ones is clearly seen in Fig.~\ref{fig:HERMES(RHIC)}.
Moreover, the dependence of hadron quenching on the target atomic
number $A$ does not seem to follow any simple law
\cite{Accardi:2005mm,Gallmeister:2007an,Airapetian:2007vu}.
For these reasons, it is not possible to directly use the results of 
Fig.~\ref{fig:coldquenchexp} to estimate cold nuclear matter
effects in h+A collisions, but we need to resort to
model computations. There exist 2 classes of models
which can reproduce nDIS data: (i) energy loss models
\cite{Wang:2002ri,Guo:2000nz,Wang:2001if,
Arleo:2003jz,Arleo:2002kh,Accardi:2005mm}, which
assume that partons hadronize well outside the target
nucleus, and loose energy because of gluon radiation induced by
rescatterings inside the target; (ii) prehadron absorption models
\cite{Accardi:2005mm,Accardi:2002tv,Accardi:2005jd,Kopeliovich:2003py,
Falter:2004uc,Bialas:1986cf,Gallmeister:2007an}, which
assume that a colorless prehadron is produced inside the target
and can be ``absorbed'' via inelastic scatterings on the nucleons.
As already remarked, the order of magnitude of the parton lifetime has
not yet been experimentally or theoretically established, and both
classes of models remain viable \cite{Accardi:2006ea}.
Hadron production in $h+A$ collisions has a large contribution from
gluon fragmentation, but this process has not been incorporated in
absorption models, so far. Therefore I chose to use energy loss models
for our estimate. In particular, I will use the BDMS framework as
implemented by Salgado and Wiedemann in
\cite{Salgado:2003gb,Salgado:2002cd} and applied to nDIS in 
\cite{Arleo:2003jz,Accardi:2005mm}. In this model, the nucleus is 
considered at rest. A parton, created with energy $E \approx \nu$ in
the hard 
interaction, travels through the nucleus and experiences multiple
scatterings and induced gluon bremsstrahlung. Hence, it starts
the hadronization process with a reduced energy $E-\Delta E$ where
$\Delta E$ is the energy of the radiated gluons.
The reduced quark energy at the time of
hadronization is translated into a shift of $z$ in the vacuum
fragmentation function $D$ \cite{Wang:1996yh}.
The medium modified FF is then computed as  
\begin{align} 
  \tilde D_{f/A}^h &(z,Q^2,E,E_h;\vec r) =
    \int\limits_0^{E_q} d\Delta E\; 
    p(\Delta E;\bar\omega_c,\bar R)  
 \label{eq:modFF}\\
  & \times \frac{1}{1-\Delta E/E}
    D_f^h(\frac{z}{1-\Delta E/E},Q^2) 
  + p_0(\bar R) \, D_f^h(z,Q^2) \ , \nonumber
\end{align}
where the quenching weight $\PP(\Delta E) = p(\Delta E) + p_0
\delta(\Delta E)$ \cite{Salgado:2003gb} 
is the probability distribution of an energy loss
$\Delta E$, with $p(\Delta E)$ its continuous part and $p_0$ the
probability of no energy loss. The quenching weight is computed for 
a static and uniform medium with characteristic gluon energy
$\omega_c = 0.5 \hat q L^2$ 
and size parameter $R = \omega_c L$, with $L$
the medium length and $\hat q$ the transport coefficient of the
medium, which characterizes the average transverse momentum squared
gained by the parton per unit in-medium path-length
\cite{Baier:1996sk,ArleoYellowRept}. 
However, the nucleus density is static but
non-uniform, hence the dependence of $\tilde D$ on the parton
production point $\vec r$, which on the r.h.s. is implicit in
the definition of suitable static-equivalent 
$\bar \omega_c$ and $\bar R$ \cite{Salgado:2002cd}, see
Eqs.~\eqref{eq:omegacbar}-\eqref{eq:Rbar}. They depend on a
single parameter, the transport coefficient $\hat q_0$ at the center 
of a reference nucleus.
The outlined energy-loss model can well describe light hadron
suppression in $\ell+A$ collisions at HERMES with $\hat q_0 = 0.5$
GeV$^2$/fm, fitted to $\pi^+$ production on $Kr$ targets
\cite{Accardi:2005mm,Accardi:2006ea}. I will use the same value
for computations in $h+A$ collisions. 

The mean free path for a parton in the target
nucleus is $\lambda = (\sigma\rho_A(\vec r))^{-1}$, where $\sigma$ is the
partonic cross-section and $\rho_A(\vec r)$ the nuclear
density. Assuming $\sigma$ to be independent of the atomic number, 
I can define a position-dependent transport coefficient,
\begin{align}
  \qhat_A (\vec b,y) = \frac{\qhat_0}{\rho_{0}} \rho_A(\vec b,y) \ ,
 \label{eq:qhatby}
\end{align}
where $\qhat_0 = \qhat_{\bar A}(0,0)$ is the transport coefficient at
the center of a reference nucleus of atomic number $\bar A$, and
$\rho_{0} = \rho_{\bar A}(0,0)$. 
Next, consider a parton produced at $\vec r = (\vec r_T,r_3)$ 
which propagates in the nucleus along the $r_3$ direction. 
Its average path-length $\bar L_A$ can be defined as
\begin{align}
  \bar L_A(\vec r) = 2 \frac{\int_{r_3}^\infty ds\, (s-r_3)\rho_A(\vec r_T,r_3)}
    {\int_{r_3}^\infty ds\, \rho_A(\vec r_T,r_3)} \ ,
\end{align}
and the average nuclear density $\bar\rho_A$ seen by the quark as
\begin{align}
  \bar\rho_A(\vec r) 
    = \frac{\int_{r_3}^\infty ds\, 
    \rho_A(\vec r_T,r_3)}{\bar L_A(\vec r_T,rf_3)} \ .
\end{align}
Then, from Eq.~\eqref{eq:qhatby}, the average transport coefficient
experienced by the quark can be defined as 
\begin{align}
  \bar\qhat_A (\vec r) = \frac{\qhat_0}{\rho_0} \bar\rho_A(\vec r) \ .
\end{align}
For a uniform hard-sphere of nuclear density $\rho_A(\vec r) = \rho_0
\theta(R_A-|\vec r|)$, the above definitions give $\bar L_A = R_A - r_3$,
$\bar\rho_A = \rho_0$, and $\bar\qhat_A = \qhat_0$ as it should be.  
Finally, the average characteristic gluon energy
$\bar\omega_c$ and size parameter $\bar R$ can be defined as follows:
\begin{align}
  \bar\omega_c(\vec r)& \equiv \frac{1}{2} \bar\qhat_A(\vec r) 
    \bar L_A^2(\vec r) 
    = \int_{r_3}^\infty ds\, (s-y)\qhat_A(\vec r_T,s) 
  \label{eq:omegacbar} \\
  \bar R(\vec r) & \equiv \bar\omega_c(\vec r) \bar L_A(\vec r) 
    = \frac{2 \bar\omega_C^2(\vec r)}{\int_{r_3}^\infty ds\,
    \qhat_A(\vec r_T,s)} \ , 
  \label{eq:Rbar}
\end{align}
These equations have also been used in
Ref.~\cite{Dainese:2004te,Eskola:2004cr} for computations of jet
quenching in the hot nuclear medium created in A+A collisions.
Note that they depend on only one parameter, $\qhat_0$.
We can also see that 
\begin{align}
  \bar\qhat_A(\vec r) = \frac{2}{\bar L_A^2(\vec r)} \int_{r_3}^\infty ds\,
    (s-r_3)\qhat_A(\vec r_T,s) \ ,
\end{align}
as in Ref.~\cite{Salgado:2002cd}. In that paper it was proven that one can
approximate the quenching weight for a dynamically expanding medium 
with the quenching weight for an equivalent static (and uniform) medium
characterized by the average $\bar\qhat_A$. However, the natural parameters
of the quenching weight are the gluon characteristic energy and the
size parameter. Hence, the scaling law is more properly expressed by
saying that the equivalent static and uniform medium is characterized
by the average $\bar\omega_c$ and $\bar R$ \cite{Dainese:2004te}.
For a parton propagating in a static but non-uniform medium, as in our
case, the spatial non-uniformity is equivalent to a time evolution of 
the medium. Therefore, as a rough {\it ansatz}, we may generalize the SW
scaling law to the case of the static but non-uniform medium
encountered in nDIS, and use Eqs.~\eqref{eq:omegacbar}-\eqref{eq:Rbar}
in the quenching weight evaluation. Note, however, that the 
suitability of a single parameter $\hat q$ to describe cold nuclear
matter has been recently questioned in Ref.~\cite{Vitev:2007ve}.

The parton production cross-section can be computed in LO pQCD as
discussed in Section~\ref{sec:DIS-eqNN}. Then the hadron production
cross-section including cold nuclear jet quenching can be written as
\begin{align}
  \frac{d\sigma^{pA\ra hX}}{dp_T^2dy_1dy_2} & =
    \int \frac{dz}{z^2}\,dy_1\,dy_2\, \\
  & \times \sum_{f_1}  \frac{d\hat\sigma^{pp\ra f_1X}}{dp_T^2dy_1dy_2} 
     \tilde D_{f_1/A}^h (z,Q^2,E,E_h;\vec r) \ ,
     \nonumber
\end{align}
where up to terms of order $O(M^2/s)$, the target rest frame 
parton and hadron energy are
\begin{align}
  E = p_T \cosh(y_1^*) \qquad 
  E_h = m_{hT} \cosh(y_h^*) \ ,
\end{align}
where $y^*_i = y_i + \log(\sqrt{s}/M)$ are the target rest frame
rapidities of the parton and the hadron.
Isospin corrections related to the target nucleus have been
included in the partonic cross section $d\hat\sigma^{pp\ra f_1X}$.
Finally, we can quantify cold matter final state energy loss effects
by the ratio of the above discussed cross section for collisions on 2
targets of atomic number $A$ and $B$:
\begin{align}
  R_{fs}^h(p_T,\bar y) 
    = \frac{d\sigma^{pA\ra hX}}{dp_T^2dy_1dy_2}
      \left[ \frac{d\sigma^{pB\ra hX}}{dp_T^2dy_1dy_2} \right]^{-1} \ ,
\end{align}
and the amount of hadron quenching by $1-R^h_{fs}$.

\begin{figure*}[tb]
  \centering
  \includegraphics[width=6.5cm]{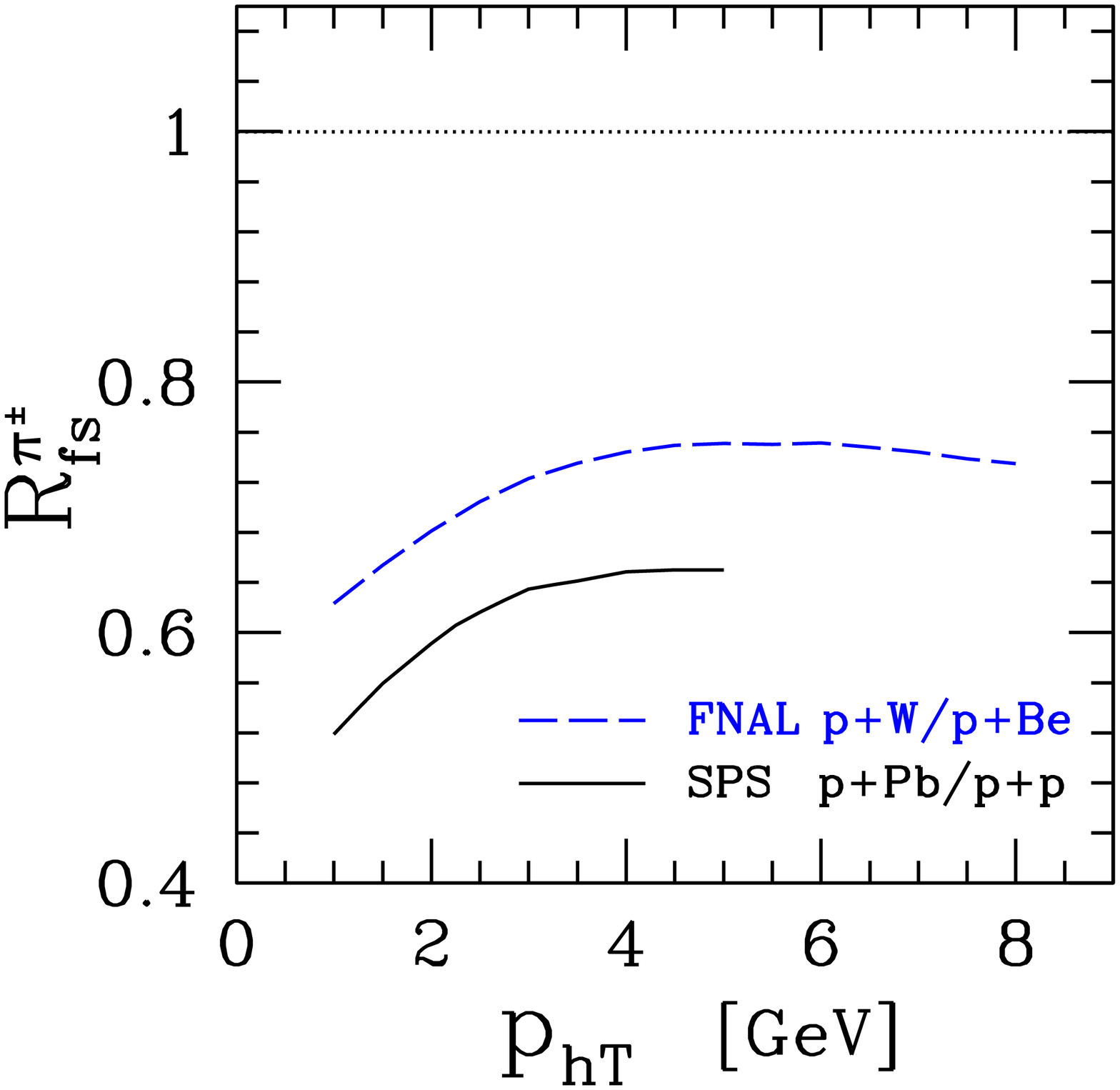}
  \includegraphics[width=6.5cm]{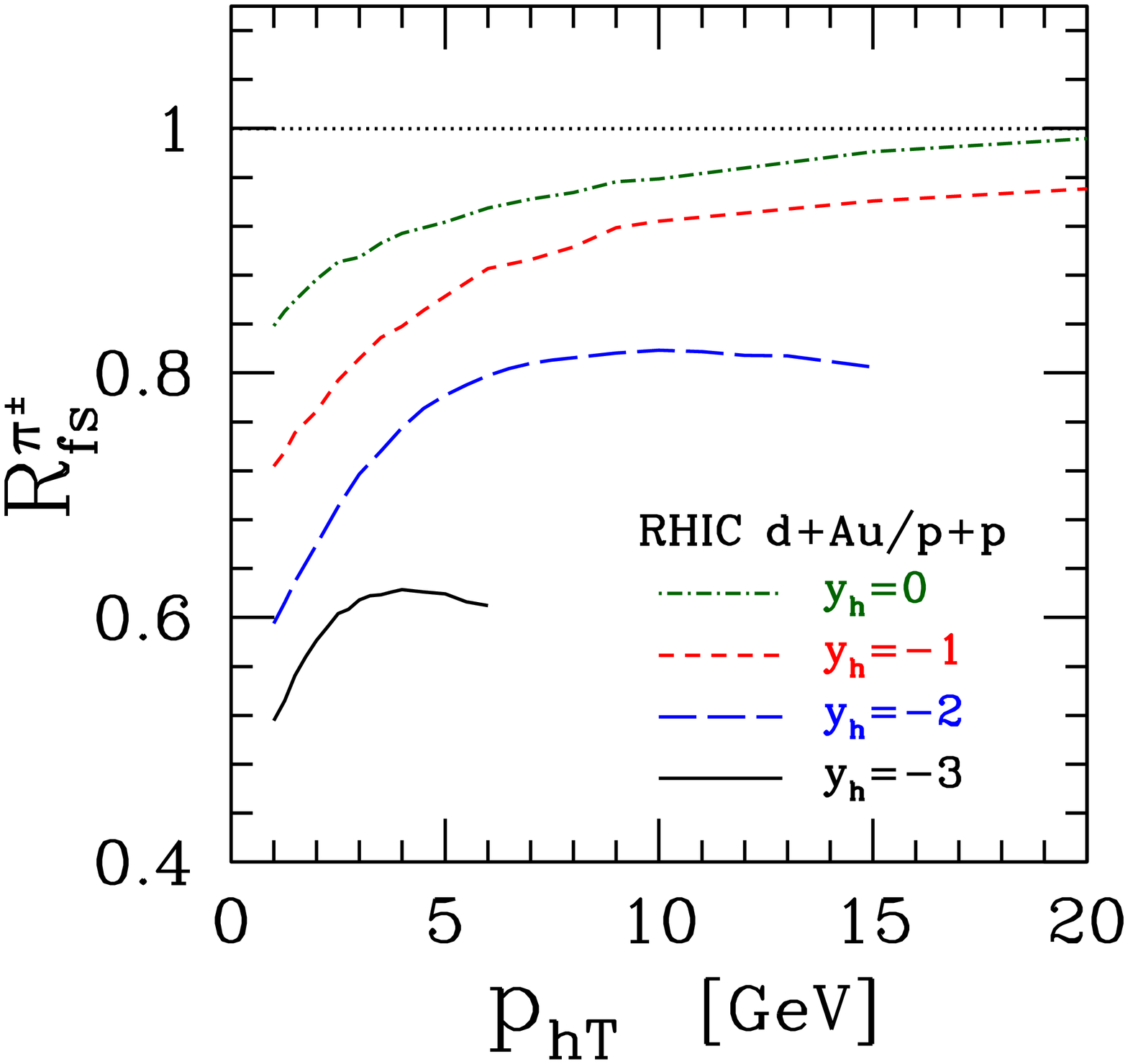}
  \caption{Energy loss model estimate of final state hadron quenching
    in cold nuclear matter
    for midrapidity pions at SPS and FNAL, and
    several negative rapidities at RHIC.}
  \label{fig:coldquenchtheory}
\end{figure*}

The computed $R^h_{fs}$ for charged pion production with no
centrality selection is presented in Fig.~\ref{fig:coldquenchtheory}. 
When examining these plots, one should keep in mind that they 
are intended only o show the extent of the suppression effects
on hadron production due to cold nuclear matter energy loss of the final
state parton. They do not include the related transverse momentum
broadening nor initial state effects, which will be commented
on in the next section. 
The plots show a substantial final state hadron quenching already
for midrapidity hadrons at SPS and FNAL energy, and for $y_h<-2$ at
RHIC. The quenching at RHIC is reduced when increasing the rapidity,
but is still non-negligible at $y_h=0$, where it is
of order 5\% at $p_T\gtrsim 10$ GeV. This may explain the small
$\pi^0$ quenching apparent in recent midrapidity PHENIX data
\cite{Adler:2006wg,Cole:2007ru}. Final state cold quenching at RHIC
should then quickly disappear at forward rapidity. At the LHC, we may
expect negligible final state effects at $y_h \gtrsim 3$ because of
the rapidity shift $\Delta y$ in Table~\ref{tab:Deltay1}.
I also found a small hadron flavor dependence at
small $p_{hT}$, not shown in the plots, showing less suppression for
kaon and proton production than for pion production.
It would be interesting to compare these estimates, obtained in the
quenching weight formalism of Salgado and Wiedemann
\cite{Salgado:2003gb}, with the results of other energy loss
implementations such as the twist-4 
formalism of Refs.~\cite{Wang:2002ri,Guo:2000nz,Wang:2001if} 
and the reaction operator
approach of Ref.~\cite{Vitev:2007ve}. A nice comparison of the
available formalisms has been recently discussed in
Ref.~\cite{Majumder:2007iu}.

\section{Initial vs. final state effects}
\label{sec:ISvsFS}

Before discussing the phenomenological relevance of the estimate of
cold nuclear matter effects obtained in the last section, we need to
discuss the importance of initial state effects, so far neglected.

The initial state parton suffers multiple scatterings and
medium-induced gluon radiation. In a simple phenomenological model
\cite{Vitev:2006bi}, the resulting energy loss may be
accounted for by a shift of the incoming parton fractional momentum,
$x_1 \ra x_1(1-\epsilon)$, with $\epsilon = \kappa A^{1/3}$ the
fractional IS energy loss.
The effect of such energy loss is felt
in a kinematic region where the flux of incoming partons varies rapidly
with $x_1$, typically at large rapidity.
Numerical estimates from \cite{Vitev:2006bi} indicate that
IS state energy loss in $d+Au$ collisions at $\sqrt{s}=19.4$ become
relevant only at forward rapidity $y-y_{cm} \gtrsim 0$. According to
the rapidity shifts listed in Table~\ref{tab:Deltay1}, 
we may expect a similar conclusion to hold for 
$y-y_{cm} \gtrsim 2 (5)$ at RHIC (LHC).

\begin{figure*}[tb]
  \centering
  \includegraphics[width=6.5cm]{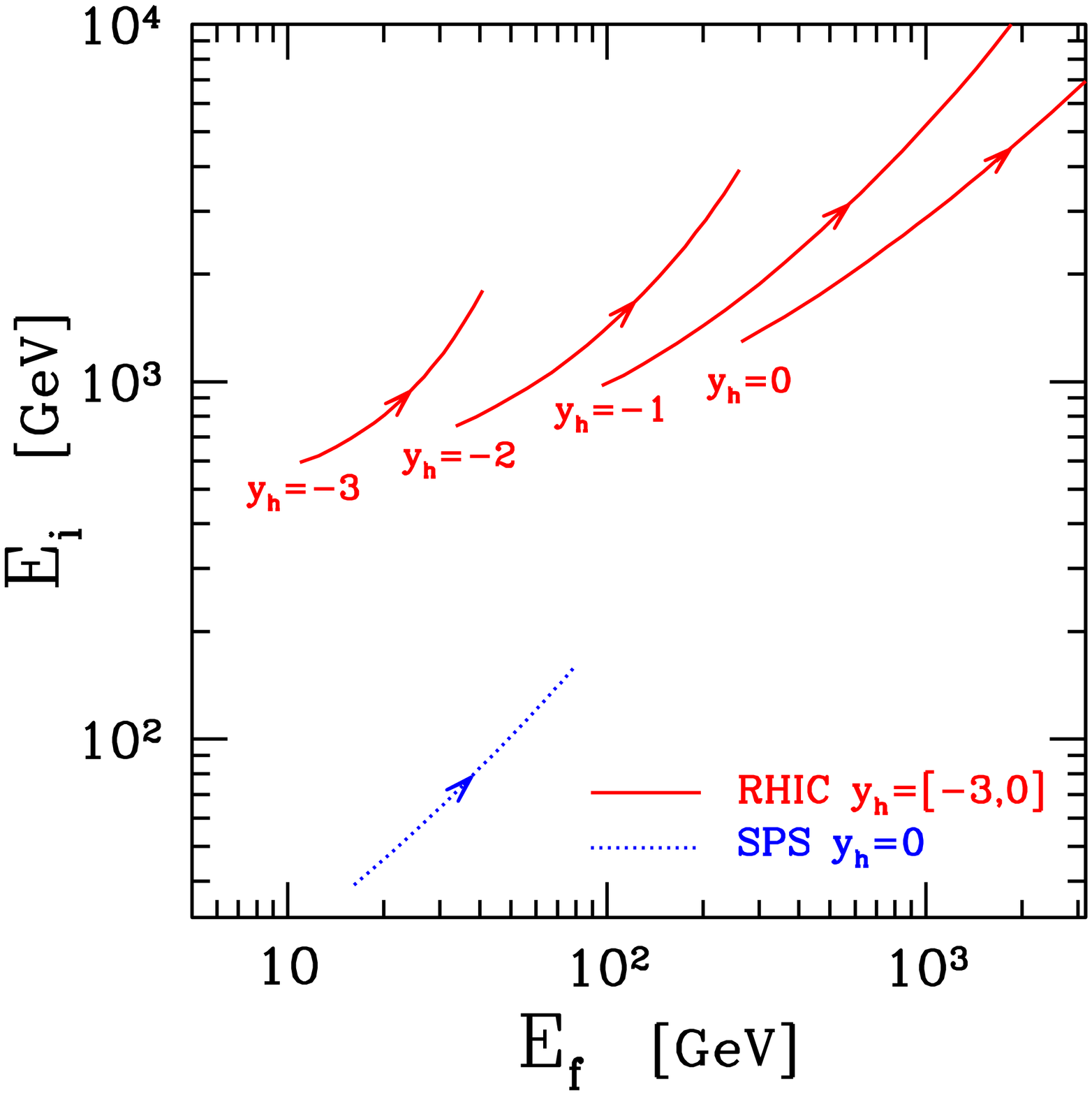}
  \includegraphics[width=6.5cm]{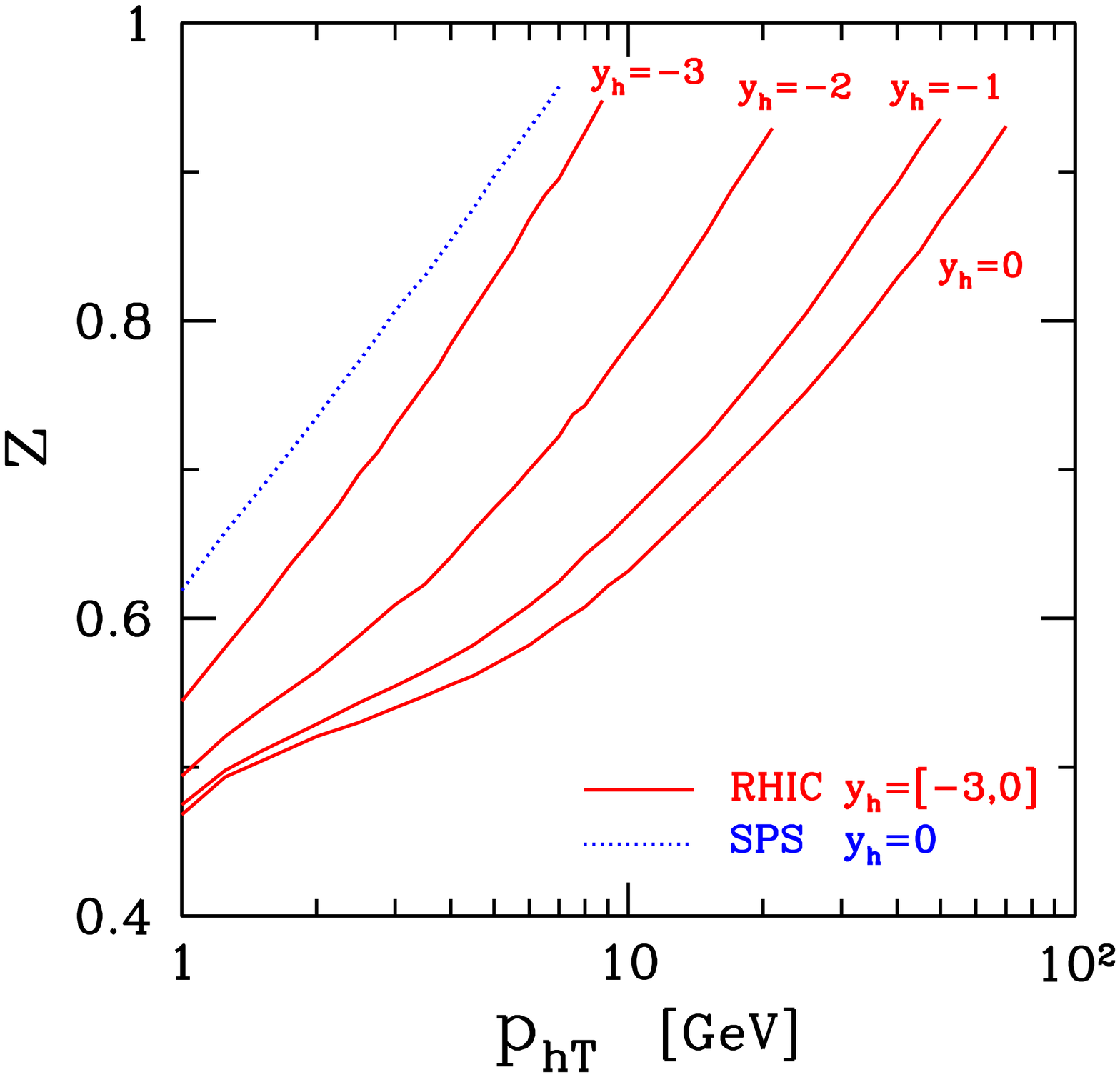}
  \caption{Correlation of initial and final state parton energy (left),
    and hadron transverse momentum and fractional momentum (right).} 
  \label{fig:EiEfphTz}
\end{figure*}

If the final state 
parton is long lived, as assumed in the theoretical estimates of
the previous section, the medium affects hadron production 
mainly through elastic and radiative energy losses. In this case, the
FS energy loss enters the computations as a shift in $z$ of the
fragmentation function, see Eq.~\eqref{eq:modFF}. Hence, differently from
IS energy loss, it is large in regions where the fragmentation
functions changes rapidly in $z$, namely at large $z$. At fixed
$p_{hT}$ the average $\vev{z}$ increases with decreasing 
rapidity and decreasing $\sqrt{s}$ (see Fig.~\ref{fig:EiEfphTz}
right). Coupling this with a decrease in final state parton energy
$E_f$ with decreasing rapidity, it is easy to explain the behavior
and large size of final state suppression shown in
Fig.~\ref{fig:coldquenchtheory}.

A consistent framework for considering the interplay of IS and FS
energy loss in the reaction operator formalism is discussed in
Ref.~\cite{Vitev:2007ve}, which presents numerical results for the
partonic fractional energy loss $\Delta E/E$ in a case
study of a homogeneous medium of fixed length $L=5$ fm. 
At any given parton energy $E$, the FS fractional energy loss
is generally smaller than the IS fractional energy loss. 
They both start at around 10\% when
$E=10$ GeV, but FS energy loss tends rapidly to 0 as $E$ increases, 
contrary to IS energy loss which stabilizes around 5\% at 
$E \gtrsim 1$ TeV. However, particle production at fixed rapidity 
in $h+A$ and $A+A$ collisions shows a strong correlation between the
IS parton energy $E_i$ and the FS parton energy $E_f$, see
Fig.~\ref{fig:EiEfphTz} left. As a result, for midrapidty hadrons at
SPS we have comparable IS and FS state energy loss of 5-10\%. For
midrapidity hadrons at RHIC, FS energy loss becomes
quite small, and IS radiation is about 5\%. For backward rapidity
production, $y_h - y_{cm} = -3$, FS energy loss is now larger than IS
energy loss, viz., 10\% vs. 5\%. A detailed computation including
realistic nuclear geometry is needed to quantify their effect on
hadron spectra.

In summary, IS and FS cold nuclear matter effects are expected to
be dominant in different rapidity regions, viz., at forward 
and backward rapidity, where the estimates presented in this paper and in
Ref.~\cite{Vitev:2006bi} indicate that they are large. 
Their effect on the midrapidity region has to be more carefully and
quantitatively considered: it depends on the center of mass energy of
the collision, and can be expected to decrease with increasing $\sqrt{s}$.   

\section{Discussion and conclusions}
\label{sec:conclusions}

In this paper, I have examined the role of final state interactions in
cold nuclear matter in modifying hadron production on nuclear targets
with leptonic or hadronic beams. Initial state parton energy loss has
been considered in \cite{Vitev:2006bi,Arleo:2002ph,Johnson:2001xf}.
Since in $\ell+A$ collisions only FS interactions are present, I built a
kinematic dictionary that 
relates the variables used for the discussion of nDIS and heavy-ion $A+B$
collisions, and demonstrated the (limited) extent to which available
experimental data on hadron suppression in nDIS can give direct 
information on final state cold nuclear matter effects in $A+B$
collisions. In this respect, the EIC
\cite{EICeAwhite,Deshpande:2005wd} will be able to efficiently 
cover the regions in phase space which the HERMES
\cite{Airapetian:2007vu} and EMC
\cite{Ashman:1991cx} experiments
could not examine. A nearly full coverage of the SPS and FNAL phase
space may alternatively 
be achieved either by a reanalysis of EMC data, or by new measurements
of hadron attenuation at the COMPASS experiment \cite{Abbon:2007pq}. 
The latter option is
particularly interesting: COMPASS has a similar kinematic coverage to
EMC, but higher luminosity and very good particle identification
capabilities. Therefore a $\mu+A$ program
at COMPASS, building on the knowledge accumulated at the HERMES
\cite{Airapetian:2007vu} and CLAS experiments
\cite{Brooks:2003cy,Hafidi:2006ig}, would greatly improve our knowledge of the
space-time evolution of hadronization, and gather vital data for the
interpretation of $h+A$ and $A+A$ collisions and the quest for the
Quark Gluon Plasma.

Hadron production in $h+A$ and $A+A$ collisions is affected by cold
nuclear matter in 2 ways. 
\begin{enumerate}
\item 
  IS and FS energy loss, and possibly FS prehadron
  absorption, suppress hadron spectra by non negligible
  amounts at forward \cite{Vitev:2006bi} and backward rapidity,
  respectively. 
\item 
  The transverse momentum broadening associated with induced radiation 
  and multiple scatterings in the medium will modify the hadron
  $p_{hT}$ spectrum, further suppressing it at $p_T \lesssim 1-2$ GeV 
  and enhancing it at intermediate momenta up to $p_T \approx  5-6$ 
  GeV \cite{Accardi:2002ik}.
\end{enumerate}

I used an energy loss model based on the BDMS formalism and tuned to
$\ell+A$ data, to estimate the size of final state cold hadron
quenching in hadronic collisions, which was
found to be large at midrapidity at fixed target SPS and FNAL energy,
and at backward rapidity at RHIC energy. It will be interesting to
compare this result with estimates based on the GLV \cite{Vitev:2007ve} and  
high-twist \cite{Wang:2002ri,Guo:2000nz,Wang:2001if} formalism for
energy loss, and on nuclear absorption models
\cite{Accardi:2005jd,Kopeliovich:2003py,Gallmeister:2007an}. 

In this paper, only the induced FS suppression has been discussed and
estimated. However, for phenomenological applications to $h+A$ and
$A+A$ collisions the interplay of suppression and enhancement of
hadron spectra will need to be elucidated. 
In $h+A$ collisions, cold nuclear 
effects will be important to understand the evolution with rapidity of
the nuclear modification factors, which decreases at forward rapidity 
\cite{Arsene:2004ux,Adams:2006uz,Back:2004bq}
but increases at negative rapidity
\cite{Abelev:2006pp,Adams:2004dv,Adler:2004eh,Abelev:2007nt}.
In $A+A$ collisions,
the time scale for the formation and equilibration of the Quark-Gluon
Plasma is much larger than the time scale for cold nuclear matter
interactions: $t_{cold} \ll t_{eq}$. Hence, both cold and hot
quenching should be considered
for a correct interpretation of experimental data. 
At RHIC midrapidity, I estimated cold quenching
of hadron spectra to be of order 5-10\%. It is much
smaller than the observed factor 4-5 hot medium quenching  
observed in central $Au+Au$
collisions, and is negligible in  first instance. At LHC cold nuclear
matter effects are likely to be negligible in a large midrapidity
interval $|y|\lesssim 3$.
At SPS energy, where
cold nuclear matter effects may be of the
same order of magnitude or larger than hot medium effects, they both need
to be taken into account in any QCD tomographic analysis to detect and
extract the properties of the Quark Gluon Plasma.

\begin{acknowledgments}
I am grateful to M.~Djordjevic, U.~Heinz, J.W.~Qiu, M.~Rosati,
G.~Sterman and I.~Vitev for valuable discussions. I would also 
like to thank P.~di~Nezza for a careful reading of the manuscript and
for drawing my attention to the possibilities offered by the COMPASS
experiment, and K.~Gallmesiter for providing me with GiBUU Monte Carlo
computations of average kinematic variables in nDIS. 
This work is partially funded
by the US Department of Energy grant DE-FG02-87ER40371.
\end{acknowledgments}

%%%%%%%%%%%%%%%%%%%%%%%%%%%%%%%%%%%%%%%%%%%%%%%%%%%%%
%%%%%%%%%%%%%%%%%%%%%  BIBLIOGRAPHY  %%%%%%%%%%%%%%%%
%%%%%%%%%%%%%%%%%%%%%%%%%%%%%%%%%%%%%%%%%%%%%%%%%%%%%

%%%%%%%%%%%%  BIBLIOGRAPHY  %%%%%%%%%%%%%%%%%%

\end{document}